\documentclass[amsmath,amssymb]{article}
\pdfoutput=1
\usepackage{authblk}
\usepackage{abstract}
\usepackage[footnotesize]{caption}
\usepackage{subcaption}
\usepackage{cite}
\usepackage{amsmath}
\usepackage{amssymb}
\usepackage{graphicx} 
\usepackage[latin1]{inputenc}
\usepackage{mathrsfs}
\usepackage[dvipsnames]{xcolor}
\usepackage{hyperref}
\usepackage[hmarginratio=1:1,top=32mm,columnsep=20pt]{geometry}
\usepackage{paralist}
\usepackage{multirow}
\usepackage{multicol}
\usepackage{blindtext}
\usepackage{upgreek}

\newcommand{\bea}{\begin{eqnarray}}
\newcommand{\eea}{\end{eqnarray}}
\newcommand{\be}{\begin{equation}}
\newcommand{\ee}{\end{equation}}
\newcommand{\ba}{\begin{array}}
\newcommand{\ea}{\end{array}}

\def\gsim{\mathrel{\rlap{\lower4pt\hbox{\hskip1pt$\sim$}}
    \raise1pt\hbox{$>$}}}

\renewcommand\Affilfont{\normalsize\itshape}

\title{\fontsize{14pt}{10pt}\selectfont 
	\textbf{Lepton-Trijet and Displaced Vertex Searches for Heavy Neutrinos at Future Electron-Proton Colliders }
	}
\author[1]{Stefan~Antusch\thanks{\texttt{stefan.antusch@unibas.ch}}}
\author[2]{Oliver~Fischer\thanks{\texttt{oliver.fischer@kit.edu}}}

\author[1]{A.~Hammad\thanks{\texttt{ahmed.hammad@unibas.ch}}}

\affil[1]{\Affilfont Department of Physics, University of Basel, \authorcr 
 		  \Affilfont Klingelbergstr.\ 82, CH-4056 Basel, Switzerland
 		   \authorcr\mbox{}}
\affil[2]{\Affilfont Institute for Nuclear Physics, Karlsruhe Institute of Technology
           \authorcr 
 		  \Affilfont  Hermann-von-Helmholtz-Platz 1, D-76344 Eggenstein-Leopoldshafen, Germany
 		   \authorcr\mbox{}} 		   
\date{}

\begin{document}
\maketitle

\begin{abstract}
\noindent Electron proton (ep) colliders could provide particle collisions at TeV energies with large data rates while maintaining the clean and pile~up-free environment of lepton colliders, which makes them very attractive for heavy neutrino searches. Heavy (mainly sterile) neutrinos with masses around the electroweak scale are proposed in low scale seesaw models for neutrino mass generation. In this paper, we analyse two of the most promising signatures of heavy neutrinos at ep colliders, the lepton-flavour violating (LFV) lepton-trijet signature and the displaced vertex signature. In the considered benchmark model, we find that for heavy neutrino masses around a few hundred GeV, the LFV lepton-trijet signature at ep colliders yields the best sensitivity of all currently discussed heavy neutrino signatures (analysed at the reconstructed level) up to now.
\end{abstract}

\section{Introduction}
The observation of neutrino flavour oscillations implies that the neutrino degrees of freedom of the Standard Model (SM) of elementary particles are not massless. At least two of them must obtain a small mass in order to explain the experimental results. Generating such masses requires physics beyond the SM, either an extended Higgs sector or the addition of extra fermions with neutral components. When these fermions are simply introduced as total singlets under the gauge group of the SM, they are often referred to as ``sterile'' neutrinos. 

In addition to a Majorana mass term, the singlet fermions can also have a Yukawa-type interaction which couples them to the SM neutrinos contained in the lepton SU(2)$_\mathrm{L}$-doublets and the SM Higgs doublet. After electroweak symmetry breaking, this term leads to a coupling of the neutral fermions to the SM Higgs boson as well as to a mixing in the neutral lepton mass matrix between the sterile neutrinos and the SM neutrinos. This mixing results in heavy and light mass eigenstates, the latter of which are mostly active neutrinos and observed in neutrino oscillation experiments, while the former are mostly sterile but have suppressed interactions with the weak gauge bosons. 
It is these suppressed interactions which allows for various production and decay channels of the new neutral heavy fermions and many aspects of the resulting signatures at particle colliders have been studied, see e.g.\ \cite{Antusch:2016ejd} and references therein. 

In the past, the Large Electron Positron collider (LEP) searched for heavy neutral leptons, i.e.\ ``heavy neutrinos'', and provides limits which are still very relevant, for instance via direct searches \cite{Abreu:1996pa}, or also via precision tests \cite{Antusch:2014woa}.
Searches for heavy neutrinos at the LHC used to focus on lepton number violating signatures, see e.g.\ \cite{Antusch:2018bgr} and references therein. Recently the CMS collaboration investigated the trilepton signature \cite{Sirunyan:2018mtv}, and ATLAS the displaced vertex signature \cite{Aad:2019kiz}.
The discovery prospects for heavy neutrinos at the LHC via lepton number conserving signatures are limited due to the large backgrounds and the tiny production cross section for larger masses. Furthermore, in typical low scale seesaw models, and in benchmark models like the ``Symmetry Protected Seesaw Scenario'' (SPSS) \cite{Antusch:2015mia} which captures their main features in  a ``simplified model'', lepton number violation is not to be expected at observable rates (cf.\ figure 3 of \cite{Antusch:2017ebe}).

An interesting way to improve the prospects for discovering heavy neutrinos at the LHC may be the Large Hadron electron Collider (LHeC) \cite{Bruening:2013bga,AbelleiraFernandez:2012cc}, envisioned to be operated simultaneously, and without interference with the hadron-hadron collisions, at $\sim$1.3 TeV centre-of-mass energy and could provide a total integrated luminosity of 1 ab$^{-1}$.
It would provide valuable improvements to the PDF sets \cite{Klein:2016uwv} and thus reduce the PDF-associated systematic uncertainties, and also significantly improve some of the Higgs measurements to the subpercent level \cite{Bruning:2652313,Angal-Kalinin:2017iup}.
First discussions of searches for heavy neutrinos at an LHeC-like collider include lepton number violating signatures \cite{Ingelman:1993ve,Liang:2010gm,Blaksley:2011ey}, while ref.~\cite{Mondal:2016kof} focuses on the lepton number conserving final states including electrons.
A systematic assessment of sterile neutrino signatures at ep colliders and first sensitivity estimates in the SPSS benchmark model are given in \cite{Antusch:2016ejd}.
More generally, electron proton colliders offer unique opportunities with respect to certain Beyond the SM (BSM) physics searches, cf.\ e.g.\ \cite{Cakir:2009xi,Liang:2010gm,Zhang:2015ado,Antusch:2016ejd,Curtin:2017bxr}, see also ref.\ \cite{Azuelos:2018syu} for an overview. 
Furthermore, the Future Circular Collider (FCC) design study also includes an electron-proton collider mode, the FCC-he, which could collide the same 60 GeV electron beam from the LHeC electron linac with the 50 TeV proton beam from the FCC-hh, giving rise to a centre of mass energy of about 3.5 TeV \cite{Mangano:2018mur,Benedikt:2018csr}.

In this article we study in depth two of the most promising direct search channels for sterile neutrinos at ep colliders, based e.g.\ on the sensitivity estimates in ref.\ \cite{Antusch:2016ejd}. In section 2 we recapitulate the model, and in section 3 we analyse the prospects for the lepton flavor violating lepton-trijet signature at the reconstructed level including the dominant backgrounds, and we carry out an improved analysis for the displaced vertex searches with the full detector geometry and event kinematics.
In section 4 we summarize our results and conclude.

\section{The model}
For our analysis, we will use the ``Symmetry Protected Seesaw Scenario'' (SPSS) benchmark model \cite{Antusch:2015mia}, which includes two sterile neutrinos with opposite charges under  a ``lepton number''-like symmetry, an extended version of the usual lepton number. The small observed neutrino masses are generated when the ``lepton number''-like symmetry is slightly broken. 
For the context of this study, we will treat the protective symmetry as being exact, which is referred to as the ``symmetry limit'' of the model. In this limit,  lepton number (LN) is conserved. When the symmetry is slightly broken (or only approximate), lepton number violation (LNV) is induced. A discussion for which parameter regions the LNV effects can be observable in the SPSS benchmark model with small symmetry breaking can be found in \cite{Antusch:2017ebe}. 

The Lagrangian density of the benchmark model, including the sterile neutrino pair $N_R^1$ and $N_R^2$ is given by:
\begin{equation}
\mathscr{L} = \mathscr{L}_\mathrm{SM} -  \overline{N_R^1}
M_N
N^{2\,c}_R - y_{\nu_{\alpha}}\overline{N_{R}^1} \widetilde \phi^\dagger \, L^\alpha
+\mathrm{H.c.}
+ \dots  \;,
\label{eqn:lagrange}
\end{equation}
where $\mathscr{L}_\mathrm{SM}$ contains the usual SM field content and with $L^\alpha$, $(\alpha=e,\mu,\tau)$, and $\phi$ being the lepton and Higgs doublets, respectively. The parameters $y_{\nu_{\alpha}}$ are the complex-valued neutrino Yukawa couplings, and $M_N$ is the sterile neutrino (Majorana) mass.
The ellipses indicate additional terms with sterile neutrinos that are decoupled from collider phenomenology as well as possible terms which slightly break the ``lepton number''-like symmetry.

Electroweak symmetry breaking yields a symmetric mass matrix of the active and sterile neutrinos, which can be diagonalized by a unitary 5 $\times$ 5 leptonic mixing matrix $U$, cf.\ \cite{Antusch:2015mia}. 
The mass eigenstates $\tilde n_j = \left(\nu_1,\nu_2,\nu_3,N_4,N_5\right)^T_j = U_{j \alpha}^{\dagger} n_\alpha$ are the three light neutrinos (which are massless in the symmetry limit) and two {\em heavy neutrinos} with degenerate mass eigenvalues $M_N$ (in the symmetry limit).
The leptonic mixing matrix governs the interactions of the heavy neutrinos, which is quantified by the active-sterile neutrino mixing angles
\begin{equation}
\theta_\alpha = \frac{y_{\nu_\alpha}^{*}}{\sqrt{2}}\frac{v_\mathrm{EW}}{M_N}\,, \qquad |\theta|^2 := \sum_{\alpha} |\theta_\alpha|^2\,,
\label{def:thetaa}
\end{equation}
with  $v_\mathrm{EW} = 246.22$ GeV being the vacuum expectation value of the Higgs field.
This allows the heavy neutrino mass eigenstates to participate in the weak current interactions, with 
\begin{eqnarray}
j_\mu^\pm & \supset &  \frac{g}{2} \, \theta_\alpha \, \bar \ell_\alpha \, \gamma_\mu P_L \left(-\mathrm{i} N_4 + N_5 \right) + \text{H.c.} \,, \label{eqn:weakcurrent1}\\
j_\mu^0 & = & \frac{g}{2\,c_W} \sum\limits_{i,j=1}^5 \vartheta_{ij} \overline{ \tilde n_i} \gamma_\mu P_L \tilde n_j\,, \\
\mathscr{L}_{\rm Yuk.} & \supset & \frac{M_N}{v_\mathrm{EW}} \sum\limits_{i=1}^3 \left(\vartheta_{i4}^* \overline{N_4^c}+ \vartheta_{i5}^*\overline{N^c_5}\right) h\, \nu_i +\text{ H.c.}  \,,
\label{eqn:weakcurrent2}
\end{eqnarray}
and where $g$ is the weak coupling constant, $c_W$ the cosine of the Weinberg angle, $P_L = {1 \over 2}(1-\gamma^5)$ the left-chiral projection operator, $h = \sqrt{2} \,\mbox{Re}{(\phi^0)}$ the real scalar Higgs boson and $\vartheta_{ij} :=  \sum_{\alpha=e,\mu,\tau} U^\dagger_{i\alpha}U_{\alpha j}^{}$.

In the symmetry limit of the benchmark model, only the moduli of the complex neutrino Yukawa couplings ($|y_{\nu_e}|$, $|y_{\nu_\mu}|$, $|y_{\nu_\tau}|$), or equivalently of the active-sterile mixing angles from Eq.~\eqref{def:thetaa}, ($|\theta_{e}|$, $|\theta_{\mu}|$, $|\theta_{\tau}|$), and the (w.l.o.g.\ real and positive) mass parameter $M_N$ are physical. 
Via the relation 
\begin{equation}
|V_{ \alpha N}|^2 = |\theta_\alpha|^2 \:,
\end{equation}
one can readily translate our results in terms of the neutrino mixing matrix elements $V_{ \alpha N}$ often used in the literature.

\section{Search Strategy}
\label{sec:strategy}
Electron-proton colliders provide an environment where the SM can be tested at higher centre-of-mass energies compared to electron-positron colliders, with comparably low rates of background.
In the following we consider the Large Hadron electron Collider (LHeC) \cite{AbelleiraFernandez:2012cc,Klein:2009qt,Bruening:2013bga} and the Future Circular Collider in hadron-electron collision mode (FCC-he) \cite{Zimmermann:2014qxa,Klein:2016uwv} for the search of the heavy neutrinos. 
The LHeC makes utilizes the 7-TeV proton beam of the LHC and a 60-GeV electron beam with up to 80\% polarization, to achieve a centre-of-mass energy close to $1.3$ TeV with a total of 1 ab$^{-1}$ integrated luminosity, while the FCC-he would collide the same electron beam with the 50-TeV proton beam from the FCC, resulting in the centre-of-mass energy close to $3.5$ TeV reaching 3 ab$^{-1}$ integrated luminosity.

\subsection{Heavy neutrino production at electron-proton colliders}
\begin{figure}
\begin{minipage}{0.49\textwidth}
\includegraphics[width=0.8\textwidth]{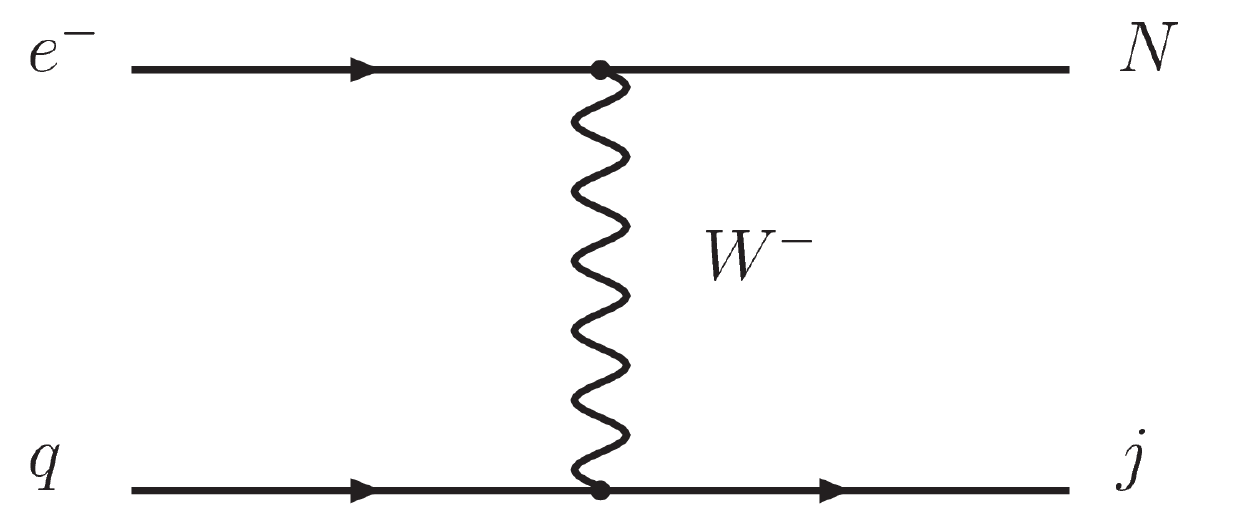}
\end{minipage}
\begin{minipage}{0.49\textwidth}
\includegraphics[width=\textwidth]{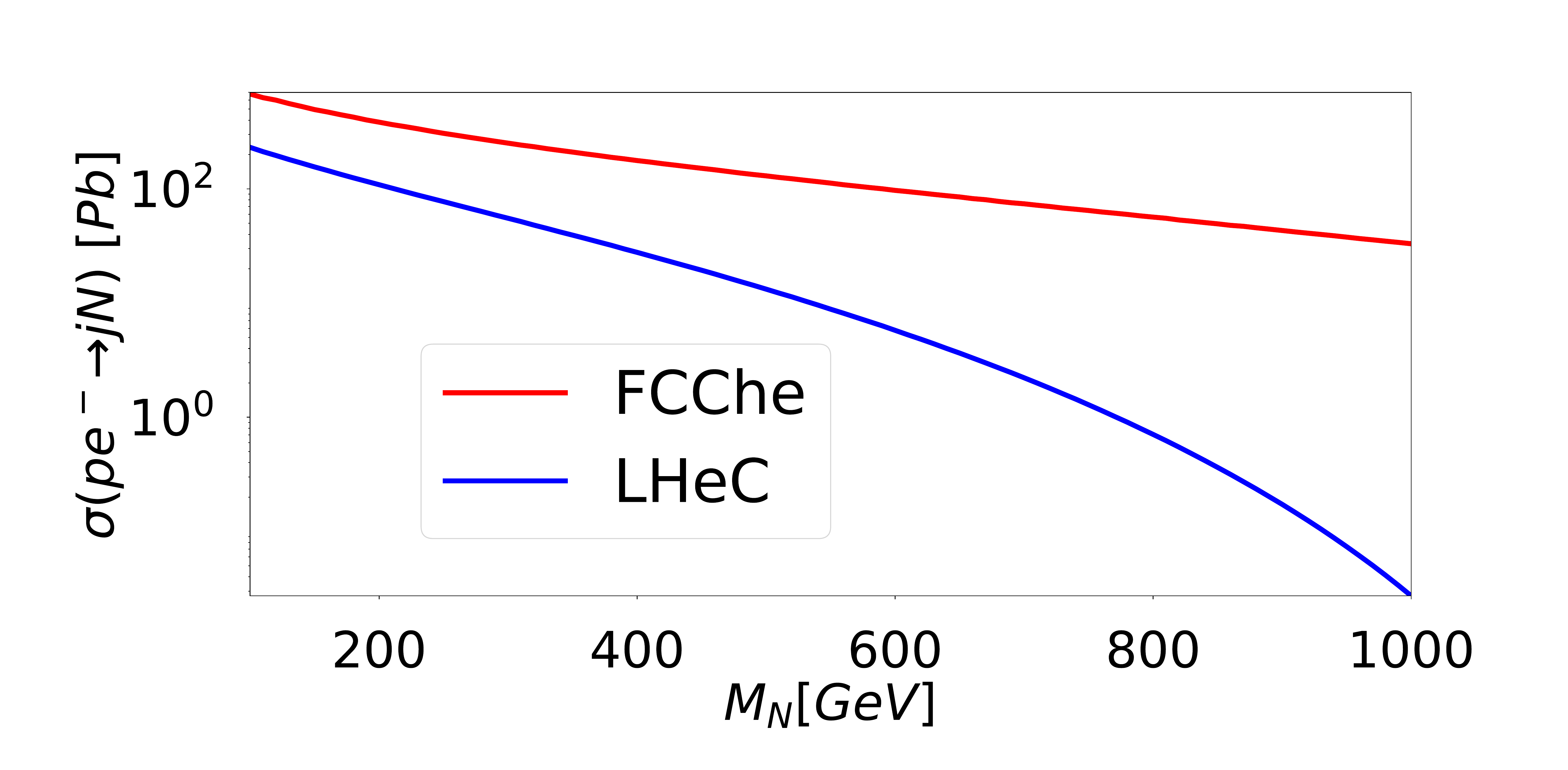}
\end{minipage}
\caption{{\it Left:} Feynman diagram representing the leading order production channel for heavy neutrinos in electron-proton scattering. {\it Right:} Cross section for heavy neutrino production in electron-proton collisions, divided by the active-sterile mixing paramter $|\theta_{e}|^2$.}
\label{fig:Nproduction-ep}
\end{figure}

At electron-proton colliders, heavy neutrinos can be produced via $t$-channel exchange of a $W$ boson together with a jet, or via $W\gamma$-fusion, which gives rise to a heavy neutrino and a $W^-$ boson. The latter channel, though suppressed by the parton distribution function of the photon within the proton, becomes increasingly important for larger centre-of-mass energies and sterile neutrino masses.
Both production channels are sensitive on the active-sterile mixing parameter $\theta_e$ only.
We show the Feynman diagram for the production mechanism via $t$-channel exchange of a $W$ boson and the production cross section in the left panel of fig.\ \ref{fig:Nproduction-ep}.

It is instructive to consider the differential cross section in the centre of mass frame, which can be expressed as:
\begin{eqnarray}
\frac{d\, \sigma}{d \, \Omega} & = & \frac{g^4 |V_{ij}|^2 |\theta_e|^2}{32 S \pi^2 ((Q^2 - m_W^2 )^2 + m_W^2 \Gamma_W^2)}  \times  \nonumber \\ 
& & \left[ \frac{S(S - M^2_N)}{4}-\frac{M^2_N(x E_p E_N + x E_p |K_N| \cos\theta)}{2} +(x E_p E_N)^2 + 2x E_p |K_N| \cos\theta \right]
\label{eq:crosssection}
\end{eqnarray}
with the squared transferred momentum $Q^2 = - M^2_N + 2 E_e (E_N - |K_N| \cos\theta)$ and the energy $S = 4 x E_e E_p$. The differential cross section depends on the energy $S$ and the two kinematic variables $Q^2$ and the Bjorken variable $x$. At electron proton colliders the Bjorken $x$ can be obtained from the measurement of the inelasticity $y_e$ as \cite{Klein:2008di}:
\begin{equation}
x = \frac{Q^2}{S y_e}\hspace{8mm} \text{with} \hspace{8mm} y_e =  1- \frac{E_N-K^z_N}{2E_e} 
\end{equation} 
with $K^z_N$ being the momentum of the scattered neutrino in $Z$-direction. The scattering angle $\theta$ is defined between the direction of the outgoing particles and the proton beam. For a large region of the parameter space with $ x \lesssim E_e/E_p$, the energy of the scattered neutrino is approximately equal to  the electron beam, which causes the cross section to peak in the negative $\theta$ direction. For more massive scattered neutrinos with $M_N \gtrsim 60$ GeV a comparatively large momentum transfer is required, which causes the heavy neutrino to scatter in the very forward direction\cite{Klein:2008di,Dainton:2006wd}.

\begin{figure}
\centering
\includegraphics[width=0.3\textwidth]{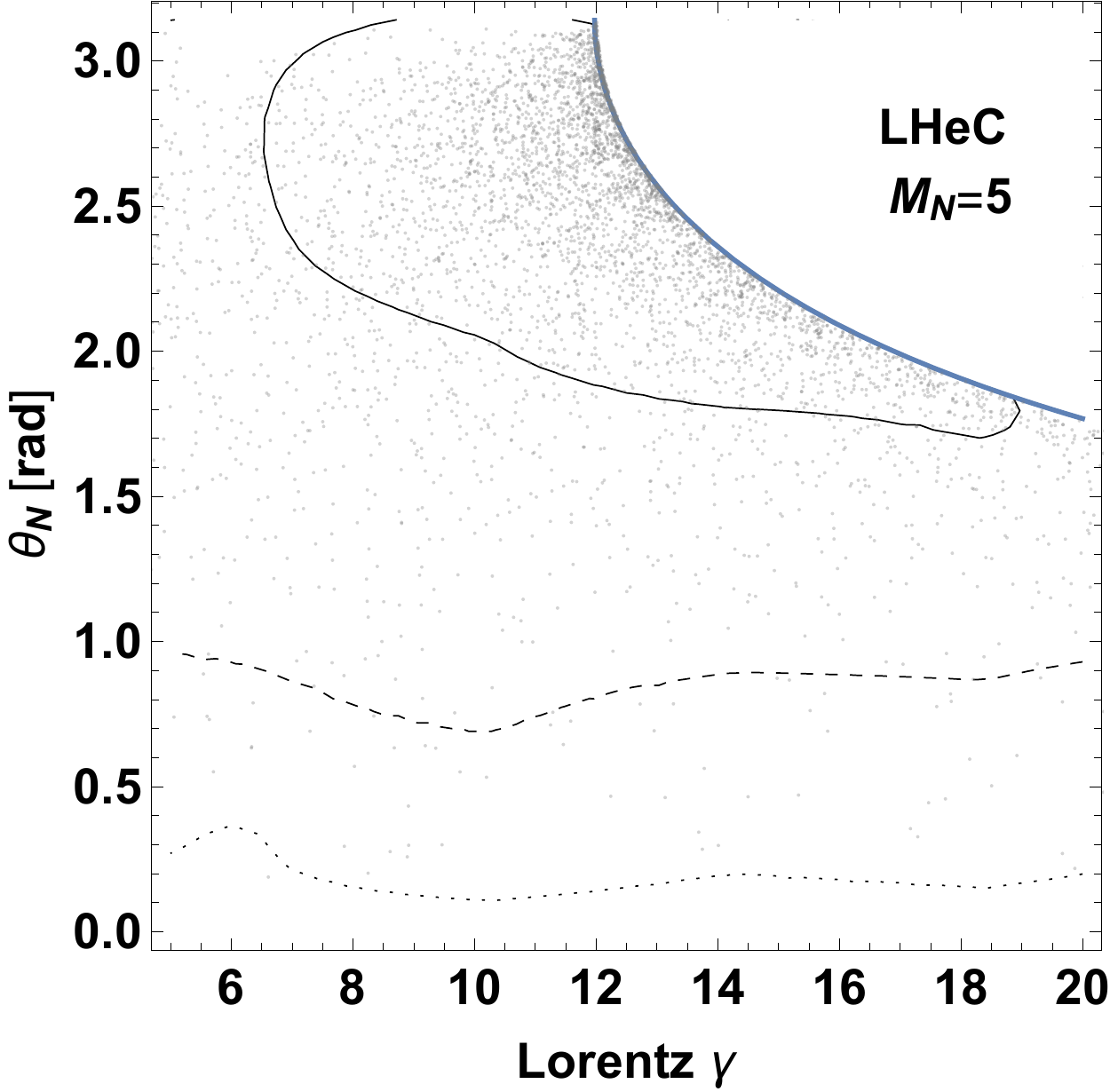}
\includegraphics[width=0.3\textwidth]{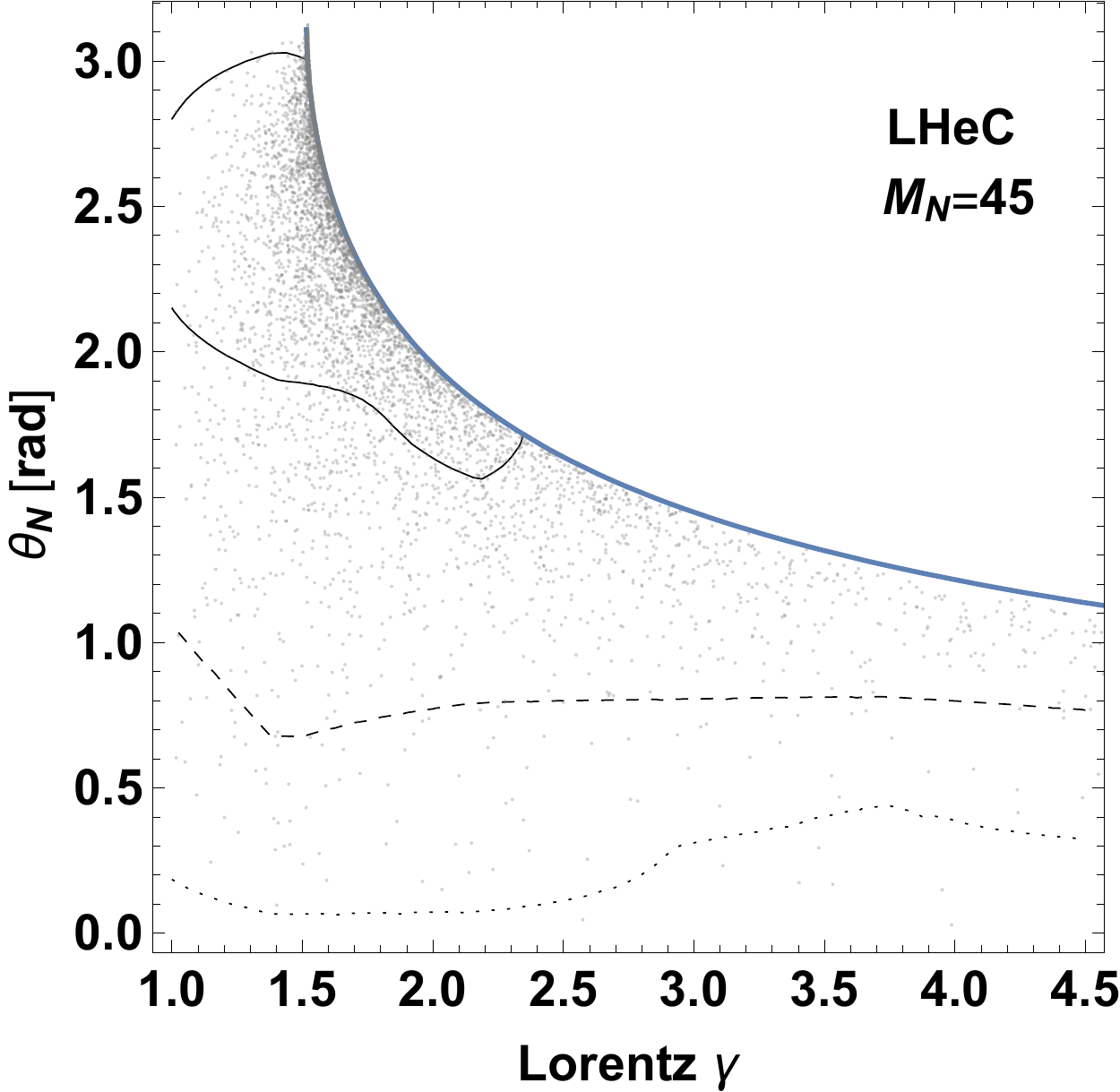}
\includegraphics[width=0.3\textwidth]{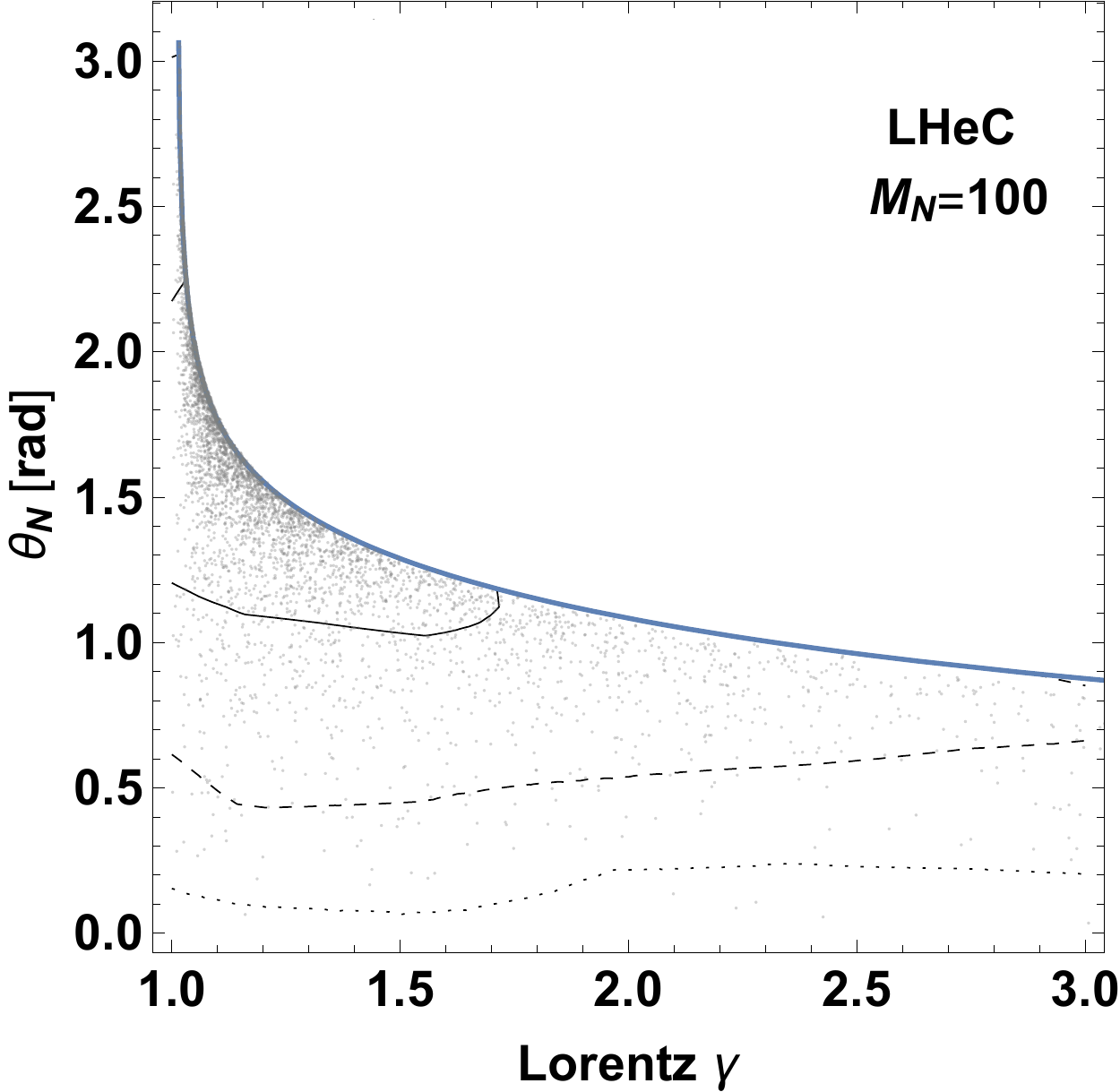}
\medskip\medskip

\includegraphics[width=0.3\textwidth]{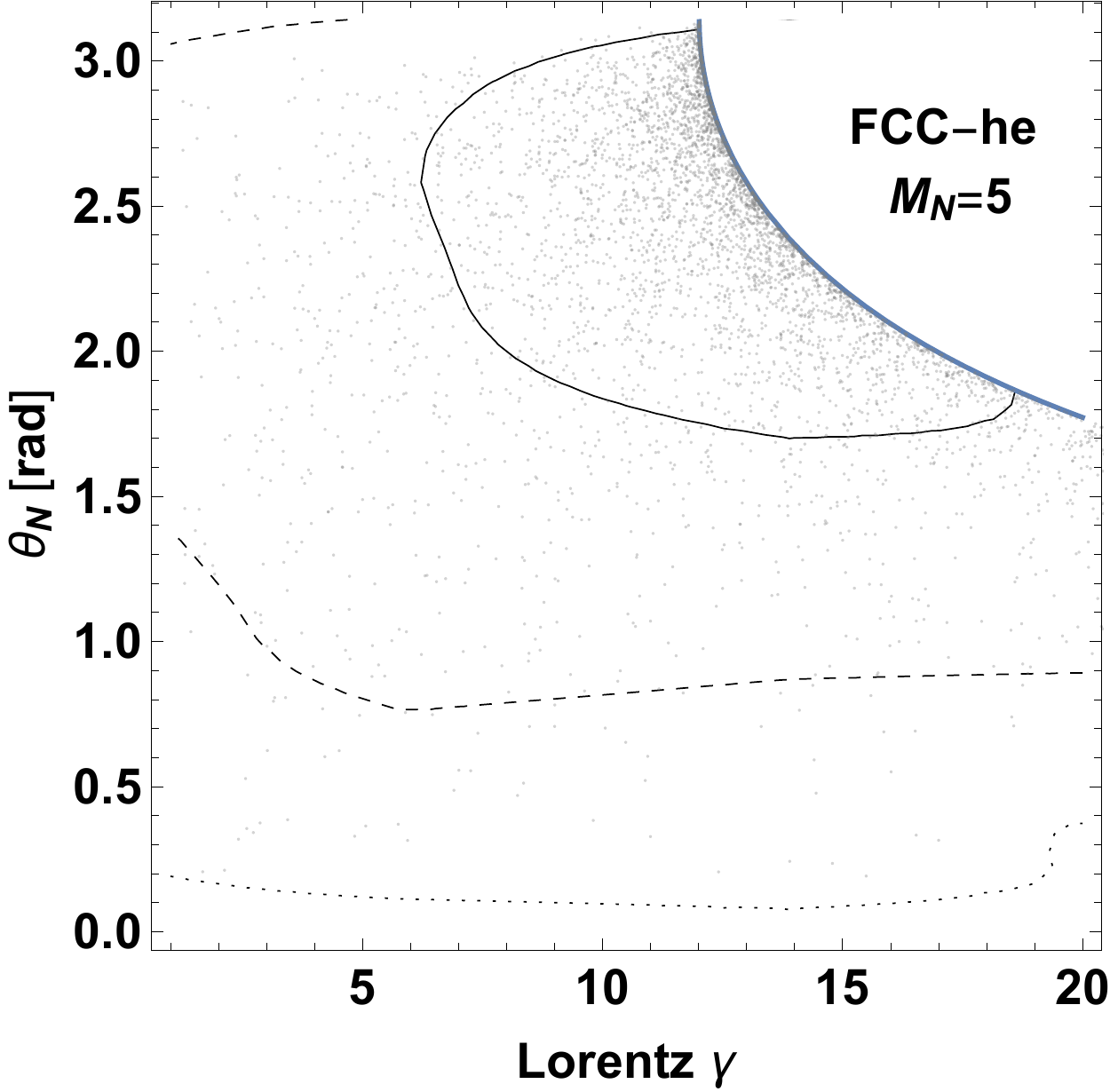}
\includegraphics[width=0.3\textwidth]{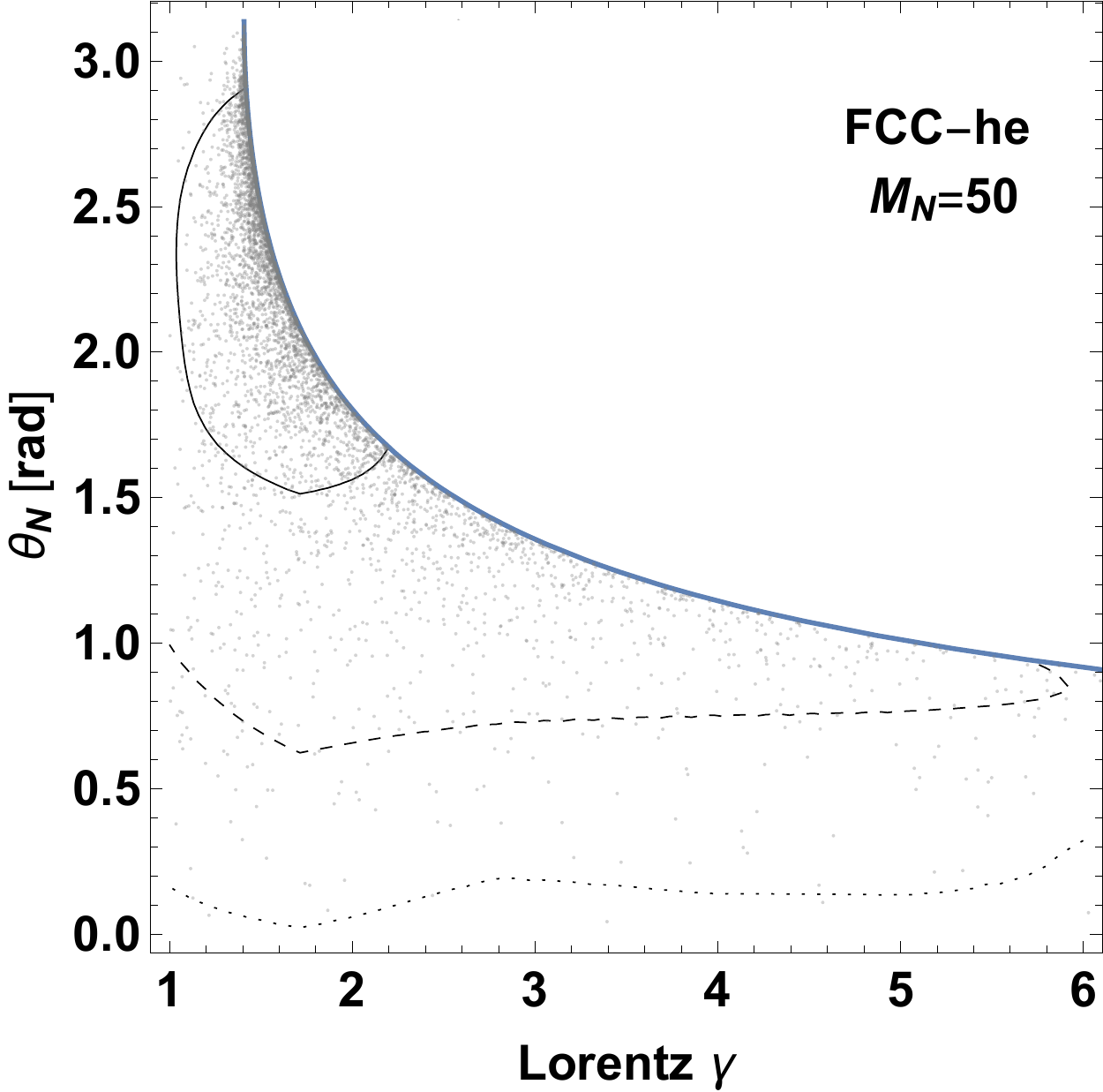}
\includegraphics[width=0.3\textwidth]{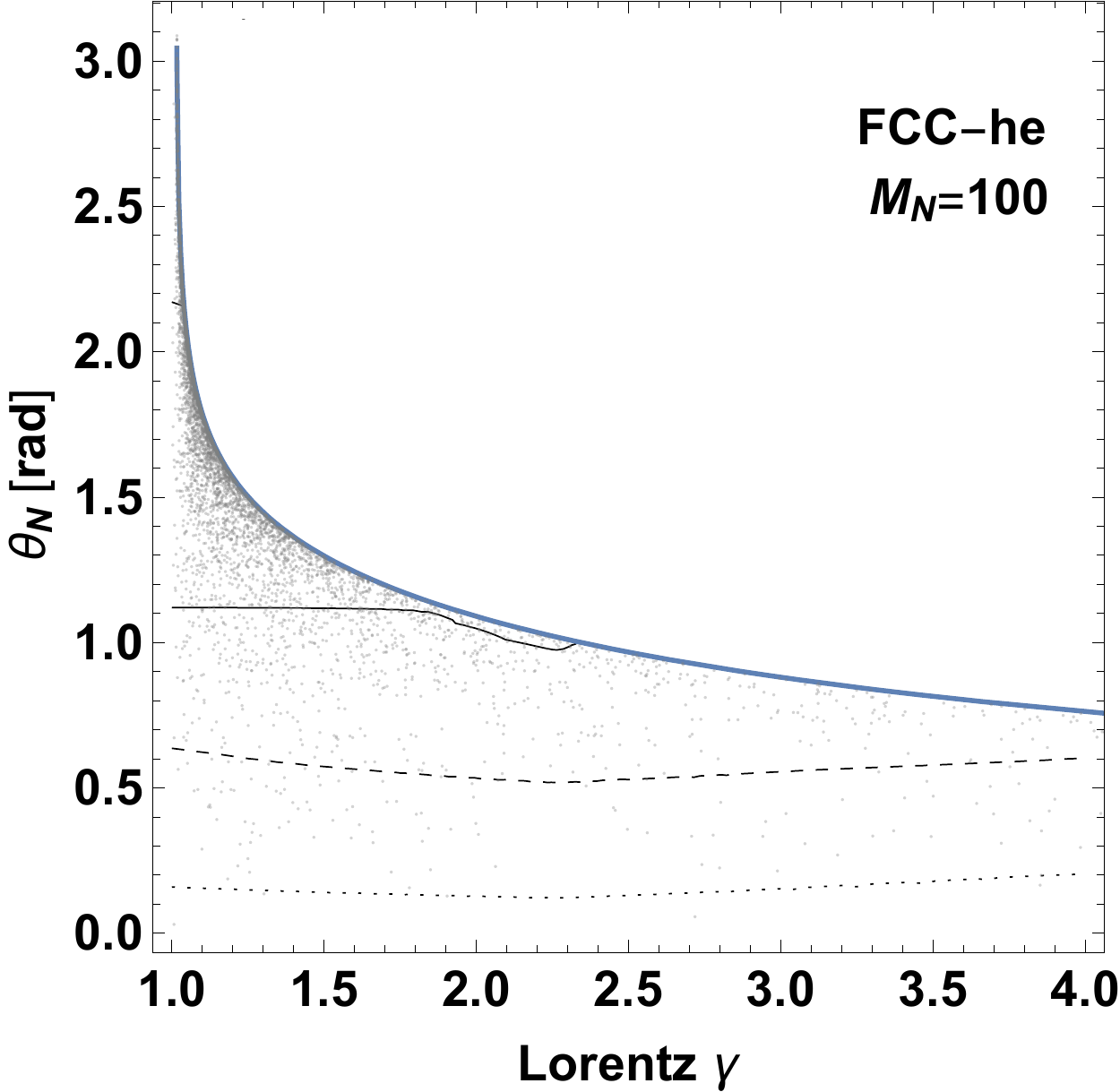}
\caption{Kinematics of the heavy neutrino produced in electron-proton collisions at the LHeC (upper row) and at the FCC-he (lower row). All masses are in GeV, the proton is in direction of $\theta = 0$. 
The plots show the distribution of the scattering angle $\theta_N$ ($10^4$ events, shown as black dots) of the heavy neutrino with respect to the beam axis versus the Lorentz boost factor $\gamma$. The black solid, dashed, and dotted line show the density contours with 68\%, 95\%, and 99\% of the points inside the contour lines.}
\label{fig:kinematics}
\end{figure}

The cross section in eq.\ \eqref{eq:crosssection} allows us to understand the kinematics of heavy neutrino production as a function of its mass, as shown in fig.\ \ref{fig:kinematics}, displayed as scattering angle of the heavy neutrino with respect to the beam axis versus the Lorentz boost factor $\gamma$. The figures were obtained from data samples with $10^4$ events and show the interpolated density contours where 68\%, 95\%, and 99\% of the points are inside the black solid, dashed, and dotted contour lines, respectively.
The correlation between the kinematical parameters $\gamma$ and $\theta$ stems from the cross section \eqref{eq:crosssection}, and can be understood from the inelasticity condition above together with the fact that $Q^2 = m_W^2$ maximises the interaction rate\footnote{Here we consider the case $M_N < m_W$. For $M_N > m_W$ this is more complicated due to suppression from the phase space versus the $W$ boson going off-shell.}:
\begin{equation}
1 - \frac{E_N - K_N^z}{2 E_e} = \frac{mW^2}{xS}\,.
\end{equation}
From this relation it follows directly, for instance, that for $\theta = \pi$ and $xS \gg mW^2$ the momentum $K_N = E_e$, while for $\theta \sim 0$ it follows that $K_N \gg E_e$.
One can identify the unphysical region for $\theta$ and $\gamma$ via $x>1$, which is shown by the black region in fig.\ \ref{fig:kinematics}.
 
We notice that the kinematics at LHeC and FCC-he produce on average similar Lorentz boosts despite the different proton beam momenta, which stems from the fact that the heavy neutrino is produced from the electron. For $M_N \leq 50$ GeV a typical Lorentz boost factor can be estimated heuristically with $E_e/M_N$. We find it interesting that the kinematical distributions are very different for the different masses $M_N$, which might allow to infer the mass of the heavy neutrino indirectly.

\subsection{Prompt Searches: Lepton-Trijets from Heavy Neutrinos}
\label{subsec:signal}

In the following we discuss the prospects of heavy neutrino detection via lepton flavor violating processes.
We consider the signal from the process chain 
\begin{equation} 
p\ e^- \to N + j  \to \mu^- + W^+ + j \to \mu^- + \, 3 j \:,
\end{equation}
where the heavy neutrino decays via the charged current into a muon and a $W^+$, which in turn decays into two jets. The branching ratio for the decay of $N$ into a muon is proportional to $|\theta_\mu|^2/|\theta|^2$, such that the rate for the process $p\ e^- \to  \mu^- + \, 3 j$ via heavy neutrinos is proportional to $|\theta_e|^2 |\theta_\mu|^2/|\theta|^2$ (cf.\ \cite{Antusch:2016ejd}).

For concreteness, we will perform our analysis assuming the relation
\begin{equation}
 |\theta_e|^2 = |\theta_\mu|^2 \gg  |\theta_\tau|^2 \: ,
\end{equation}
setting $\theta_\tau$ to zero. This allows to present them later in the results section together with the existing bounds on $\mu \to e + \gamma$, as functions of $| \theta_e \theta_\mu |$. Our results of course hold general: one simply has to replace $| \theta_e \theta_\mu |$ by $2 |\theta_e|^2 |\theta_\mu|^2/|\theta|^2$ on the y-axis of the plots showing the results for the sensitivities.

This lepton-trijet final state yields an ``unambiguous signal'' for lepton flavour violation, which means there exists no SM background process at the parton level with this final state, as discussed in \cite{Antusch:2016ejd,Arganda:2015ija}. SM backgrounds, as will be discussed below, of course exist due to possible misidentification or, e.g., from SM final states which only differ by additional light neutrinos. For the latter type of backgrounds, one expects that the kinematical distributions of the muon can be used as a powerful discriminator between signal and background. The sensitivity prospects for the LHeC and the FCC-he have been estimated in ref.~\cite{Antusch:2016ejd} at the parton level (with not optimised cuts). In this work, we will improve these sensitivity estimates.

\subsubsection{Considered Standard Model Backgrounds}
\begin{table}[h!]
\begin{center}
\begin{tabular}{|l|c|c|}
\hline
Backgrounds & $\sigma_{(LHeC)} [Pb]$ & $\sigma_{(FCChe)} [Pb]$  \\ \hline\hline
$p e^-\to j e^- VV , \quad \mbox{where}  \;VV\to jj\mu^-\mu^+$  & 0.00616  & 2.40 \\ \hline
$p e^-\to j e^- VV, \quad \mbox{where}  \;VV\to jj\mu^-\bar{\nu}_\mu$  &0.00185  & 0.45\\ \hline
$p e^-\to j \nu_e VV , \quad \mbox{where}  \; VV\to jj\mu^-\mu^+$ & 0.00606& 2.30 \\ \hline
$p e^-\to j \nu_e VV , \quad \mbox{where}  \; VV\to jj\mu^-\bar{\nu}_\mu$ &0.00180  & 0.44   \\ \hline
\end{tabular}
\end{center}
\caption{Dominant background processes considered in our analysis and their total cross sections. The samples have been produced with the following cuts: $P_T(j)\ge 5$ GeV, $P_T(l)\ge 2$ GeV and $|\eta(l/j)|\le 4.5$. }
\label{tab:signatures4}
\end{table}

The dominant SM backgrounds for the $jjj\mu^-$ signature considered in our analysis, and their total cross sections, are summarized in table \ref{tab:signatures4}. 

One very important background arises from di-vector boson production associated with jet and a neutrino, e.g $p e^-\to j \nu_e V V$ with $V= W^-,\, W^+,\, Z$. Especially when one of the $V$ is a $W^-$, decaying into $\mu^- \nu_\mu$, then the final state only differs from the signal by two additional neutrinos. Nevertheless, the light neutrino in the final state gives rise to missing energy and allows for efficient separation of this process from the signal, which comes without missing energy. 

Another important class of background comes from di-vector boson production associated with a jet and an electron, e.g $p e^-\to j e^- V V$ with $V= W^- W^+ Z$. While the signal does not have hard electrons, it contains many soft electrons due to radiative processes. Therefore one cannot simply reject events that contain electrons without decreasing the signal efficiency. For $m_N \le 200$ GeV, the distance $\Delta R(W,\mu)=\sqrt{(\Delta\eta)^2+(\Delta\phi)^2}$ between the $W^-$ and the muon is a very good discriminator, since in the background the muons always come from vector boson decays. For higher masses $m_N \ge 400$ GeV, the muons from heavy neutrino decays are highly boosted and can be distinguished from the background muons. 

The background that arises from single vector boson production with radiated jets can be reduced very well since the radiated jets are very soft and can be easily distinguished from signal jets. Also, background with single vector boson production that decays to a tau lepton pair that gives raise to $jj\mu^- + MET$ final state is highly reduced because of the missing energy and low momentum of the final state fermions that come from the tau decay. Finally, the three vector boson production is not considered since its cross section is much smaller compared to the two vector boson production processes.

\begin{figure}
\includegraphics[width=0.38\textwidth]{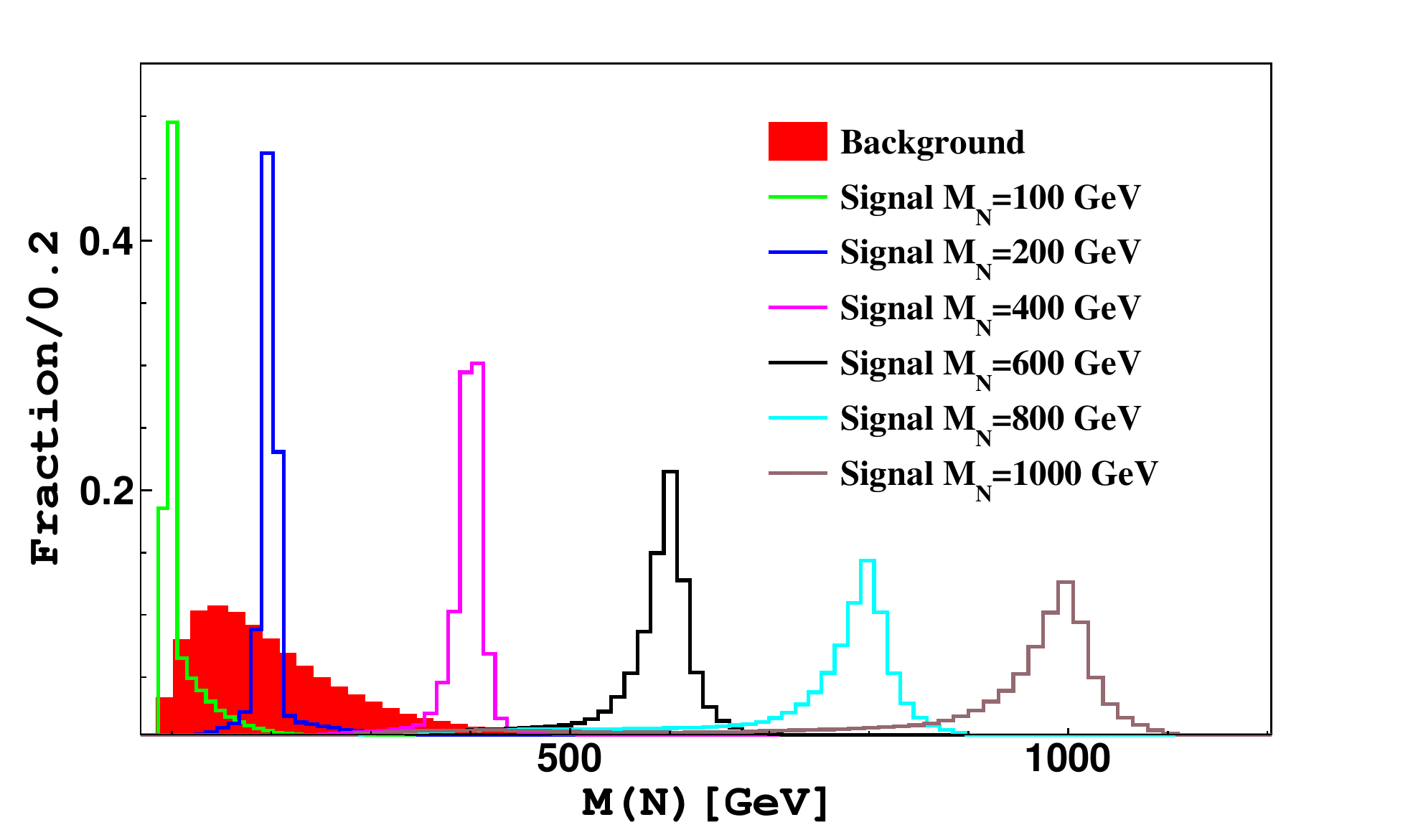}~~\includegraphics[width=0.38\textwidth]{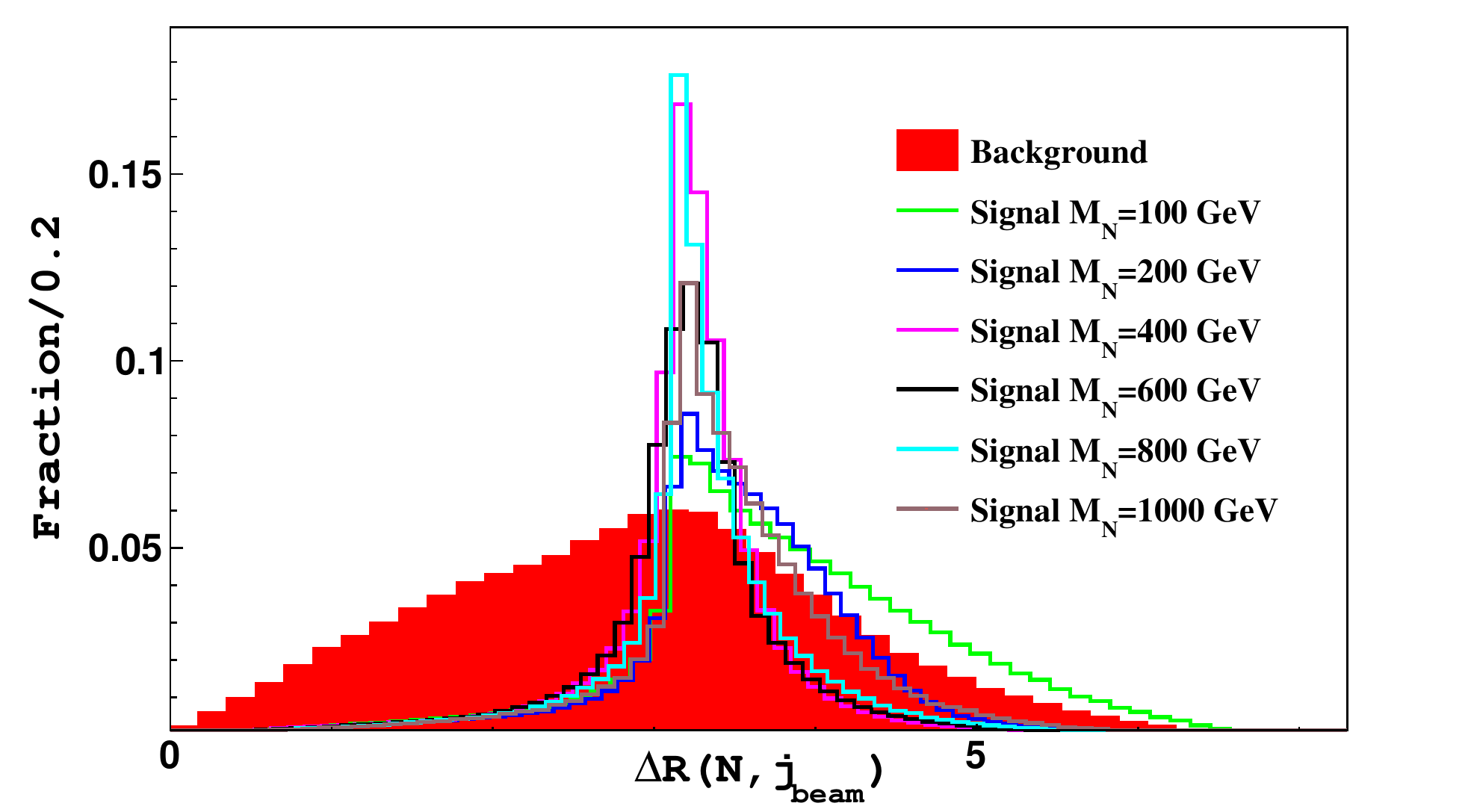}~~\includegraphics[width=0.38\textwidth]{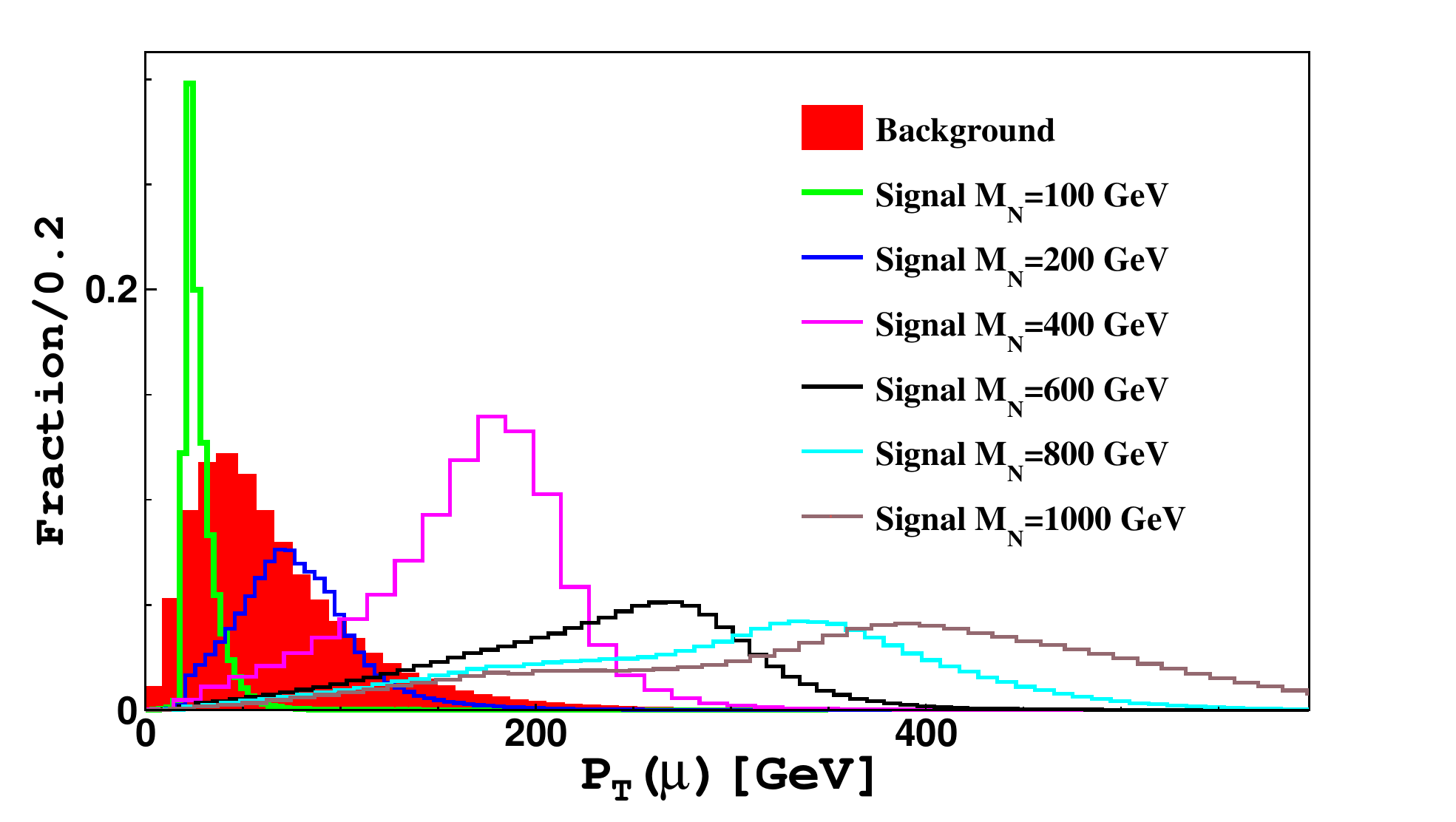}
\\
\includegraphics[width=0.4\textwidth]{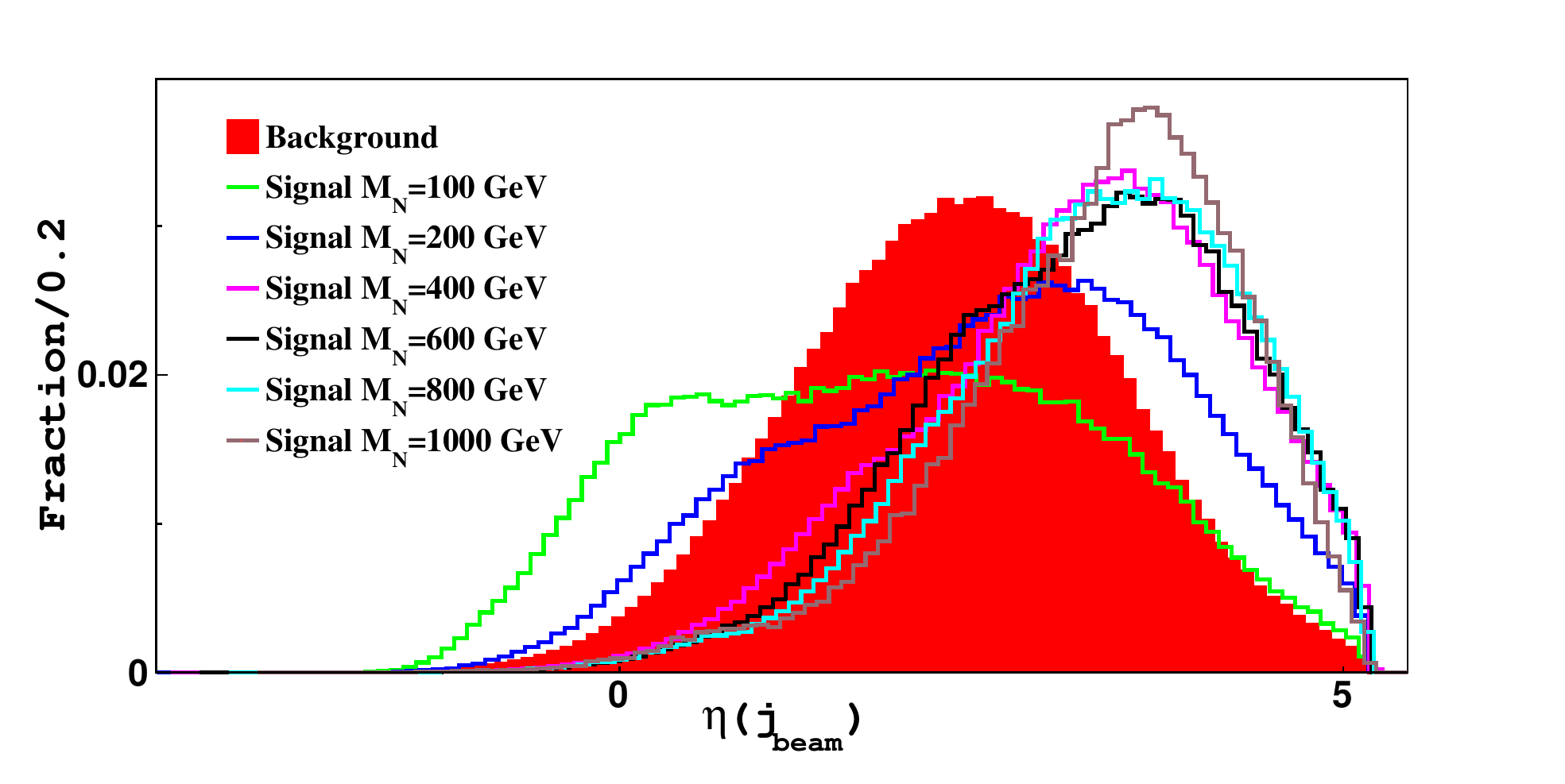}~~\includegraphics[width=0.4\textwidth]{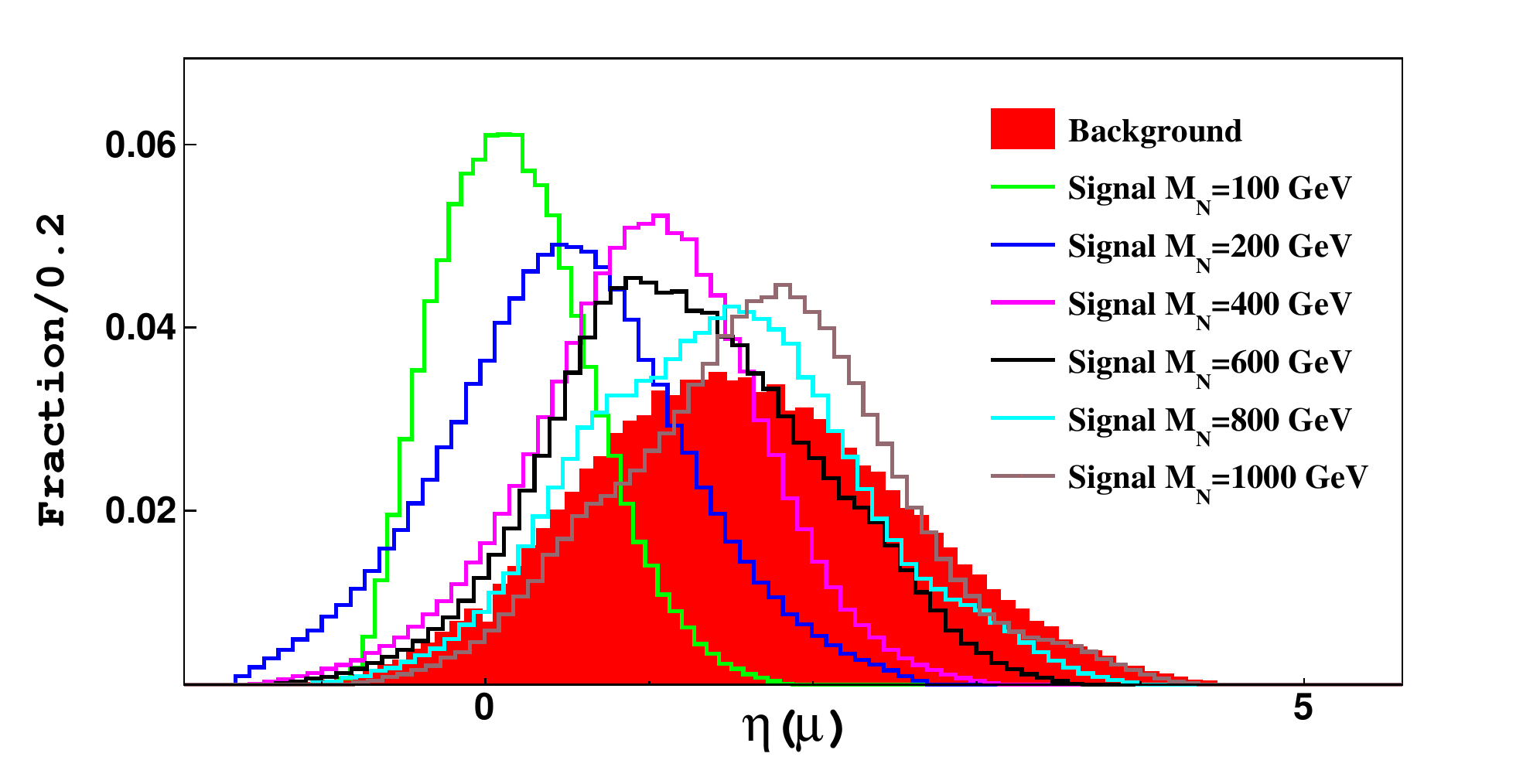}~~\includegraphics[width=0.31\textwidth]{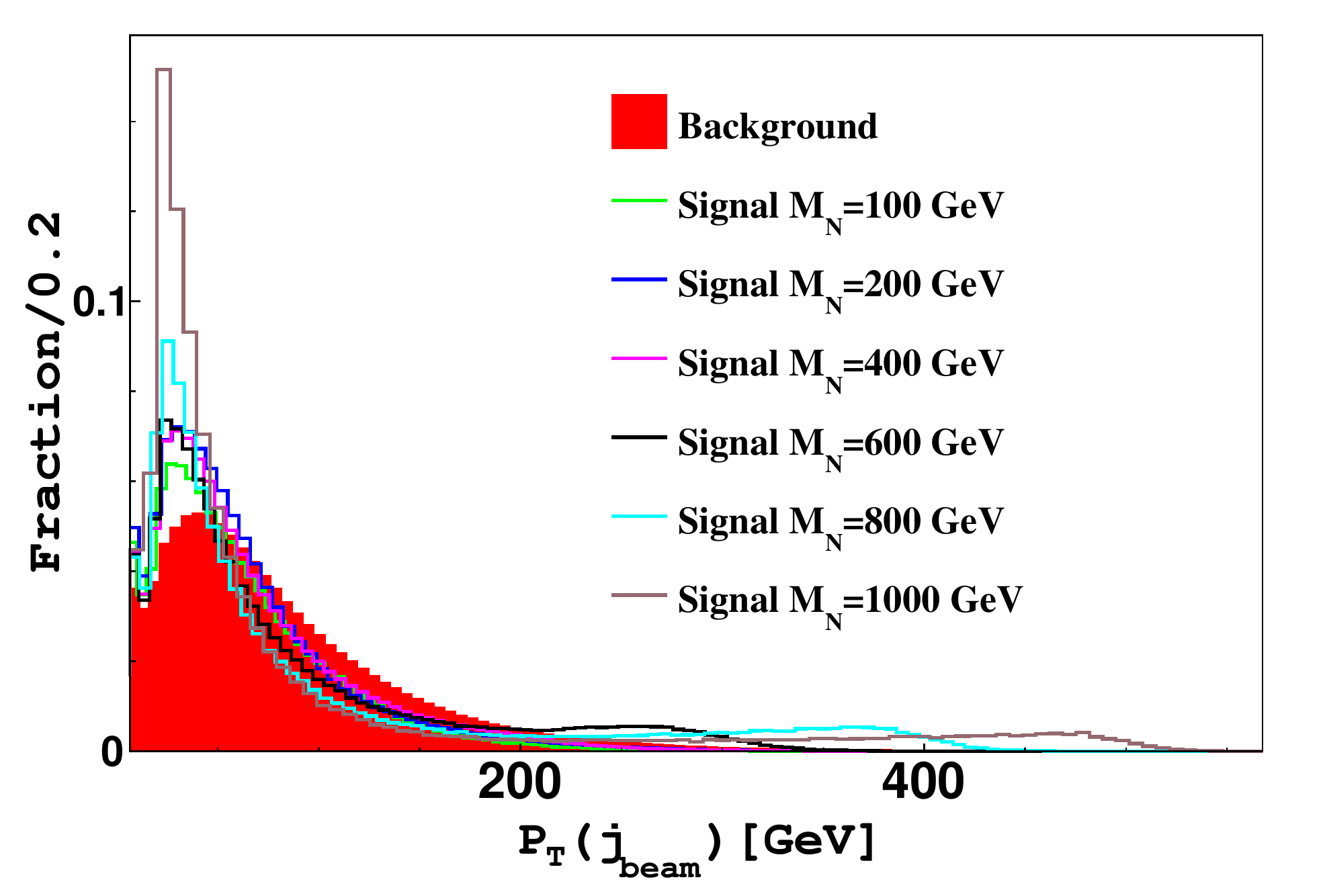}
\\
\includegraphics[width=0.37\textwidth]{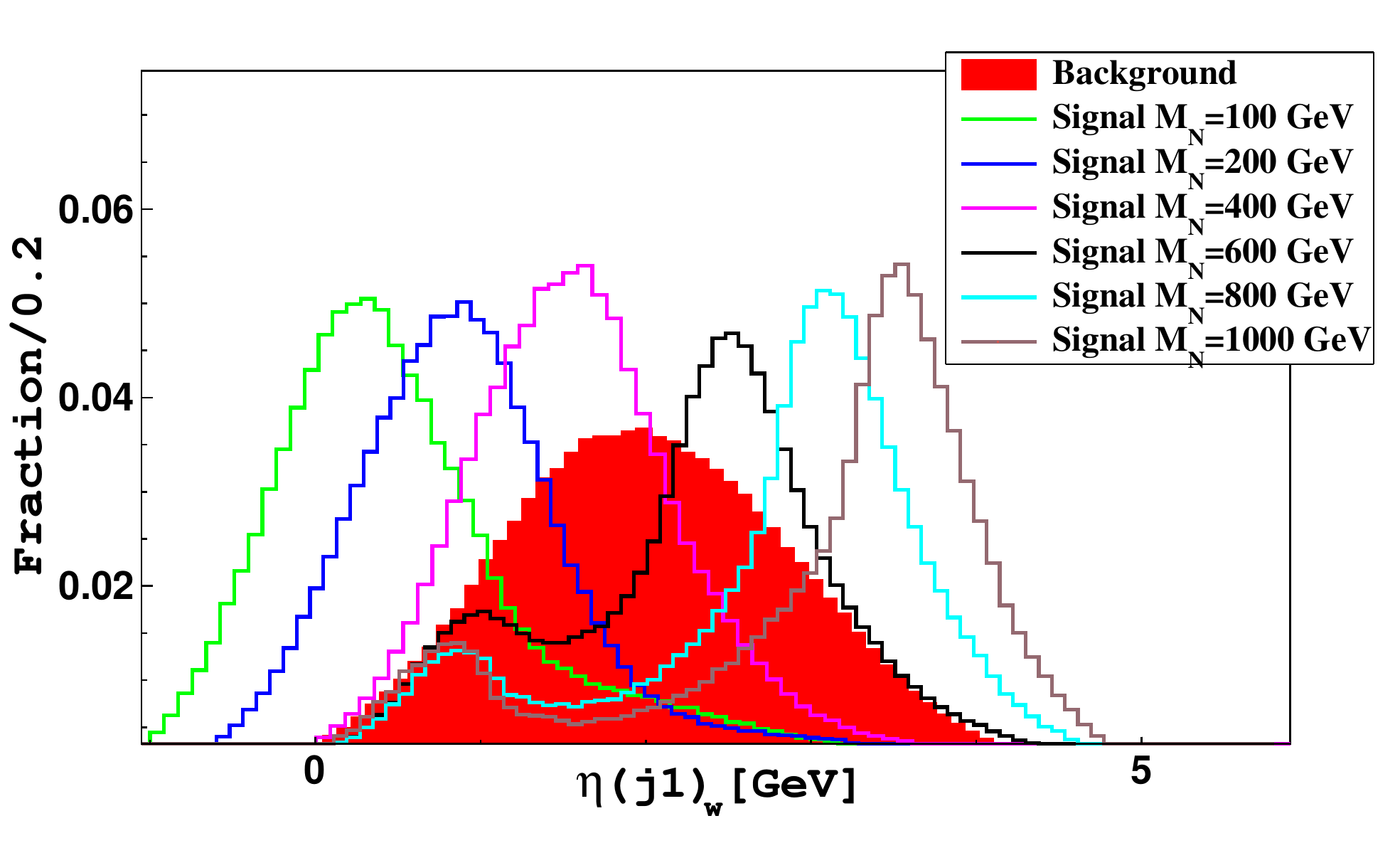}~~\includegraphics[width=0.37\textwidth]{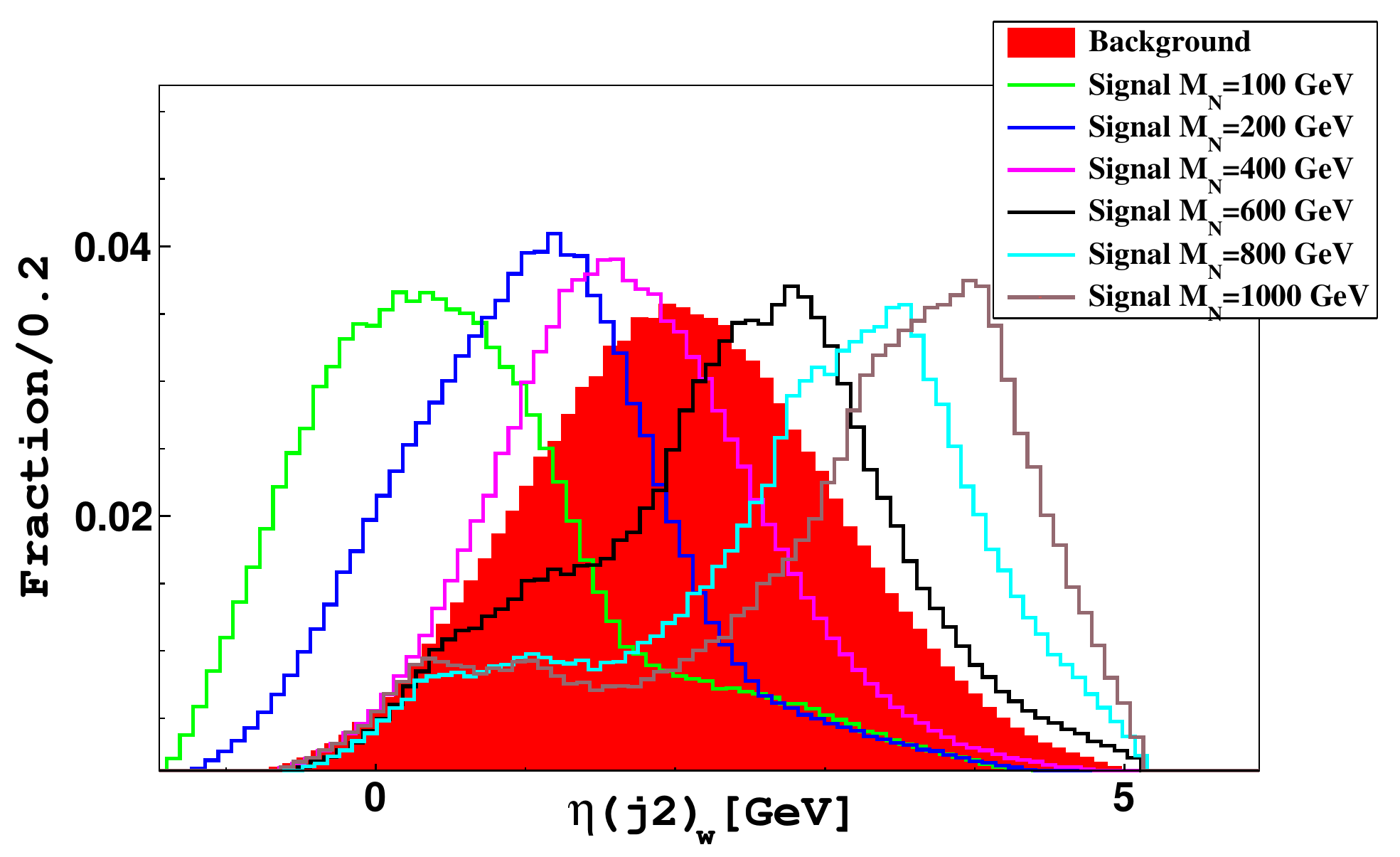}~~\includegraphics[width=0.37\textwidth]{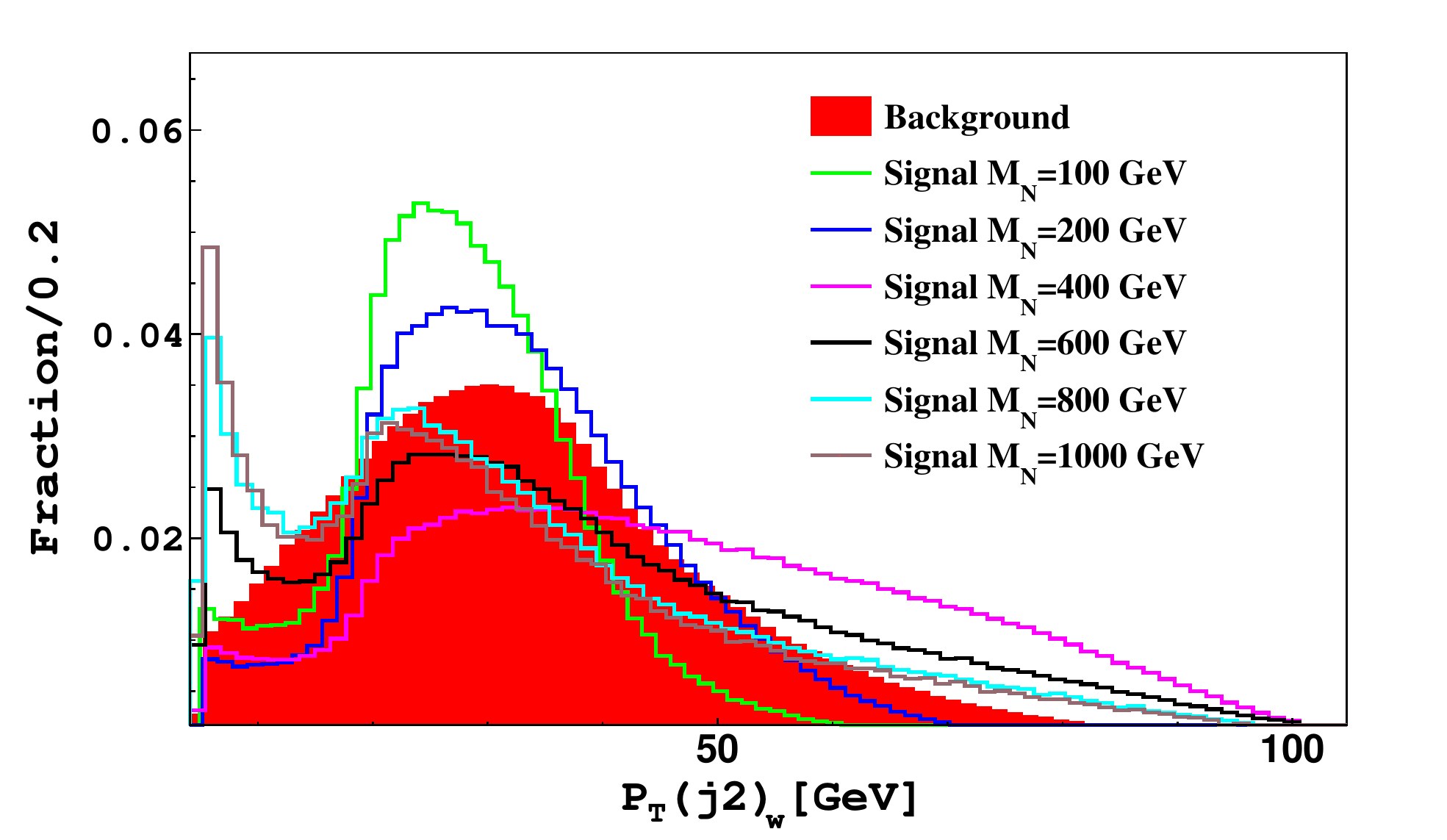}
\\
\includegraphics[width=0.35\textwidth]{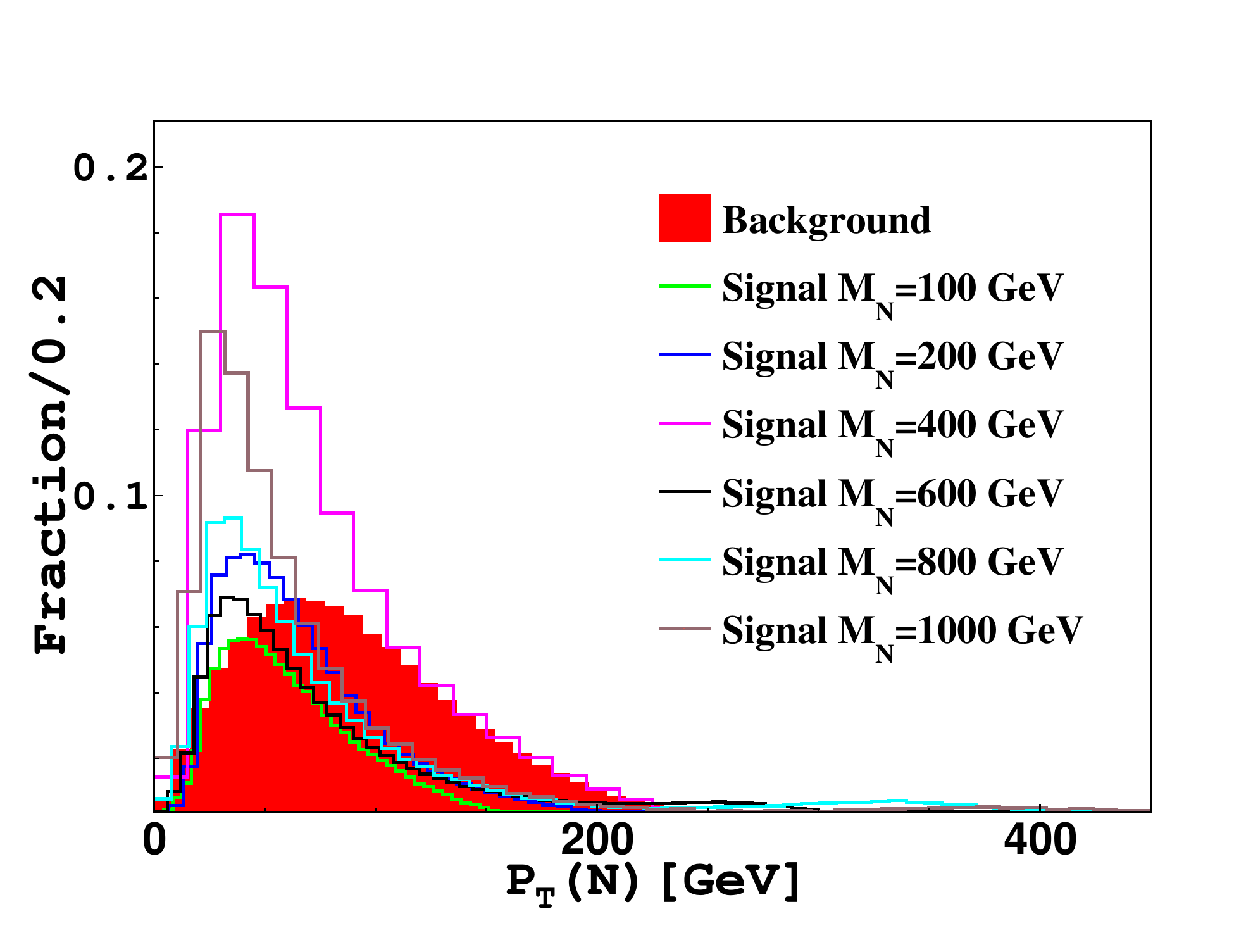}~~\includegraphics[width=0.38\textwidth]{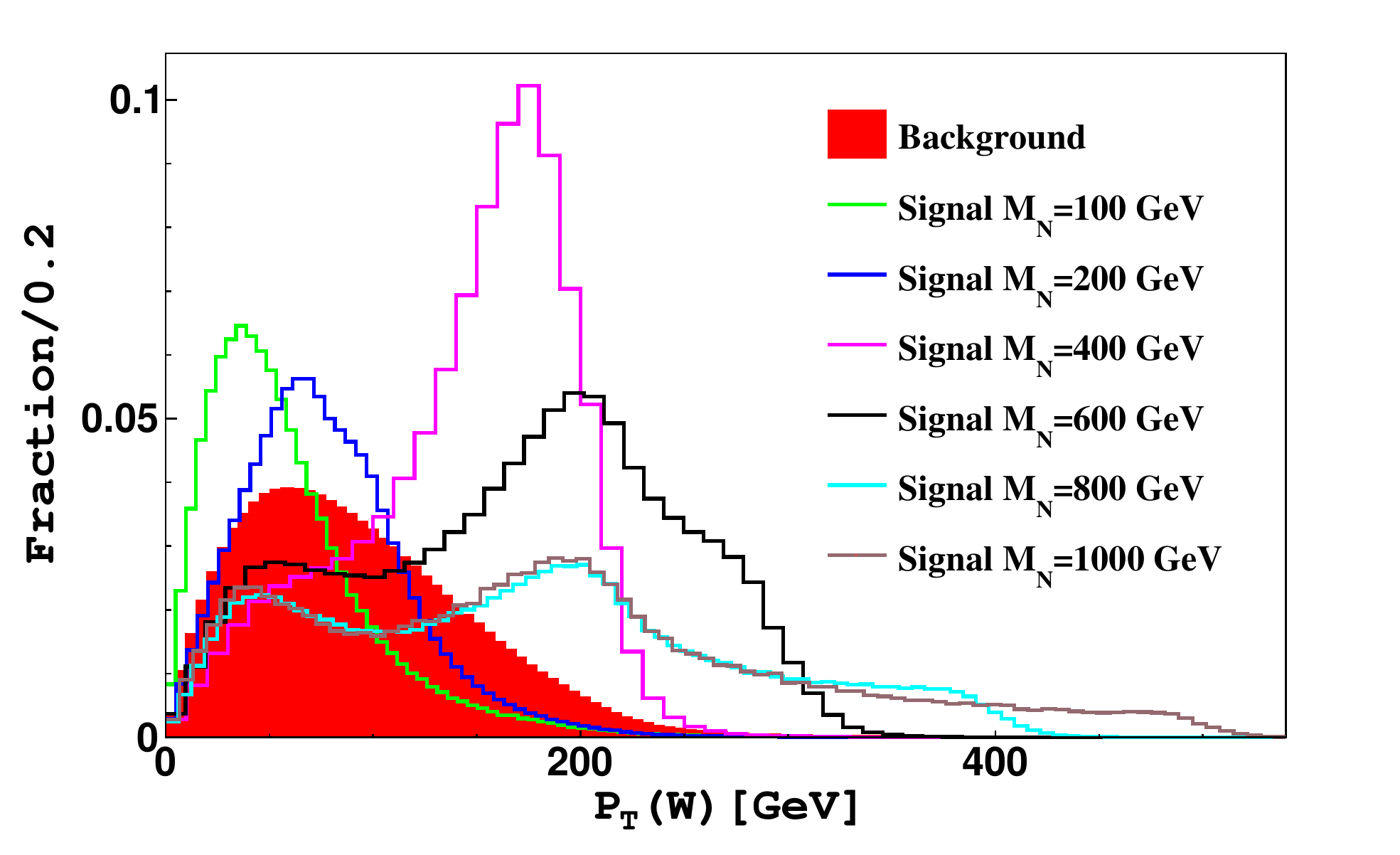}~~\includegraphics[width=0.37\textwidth]{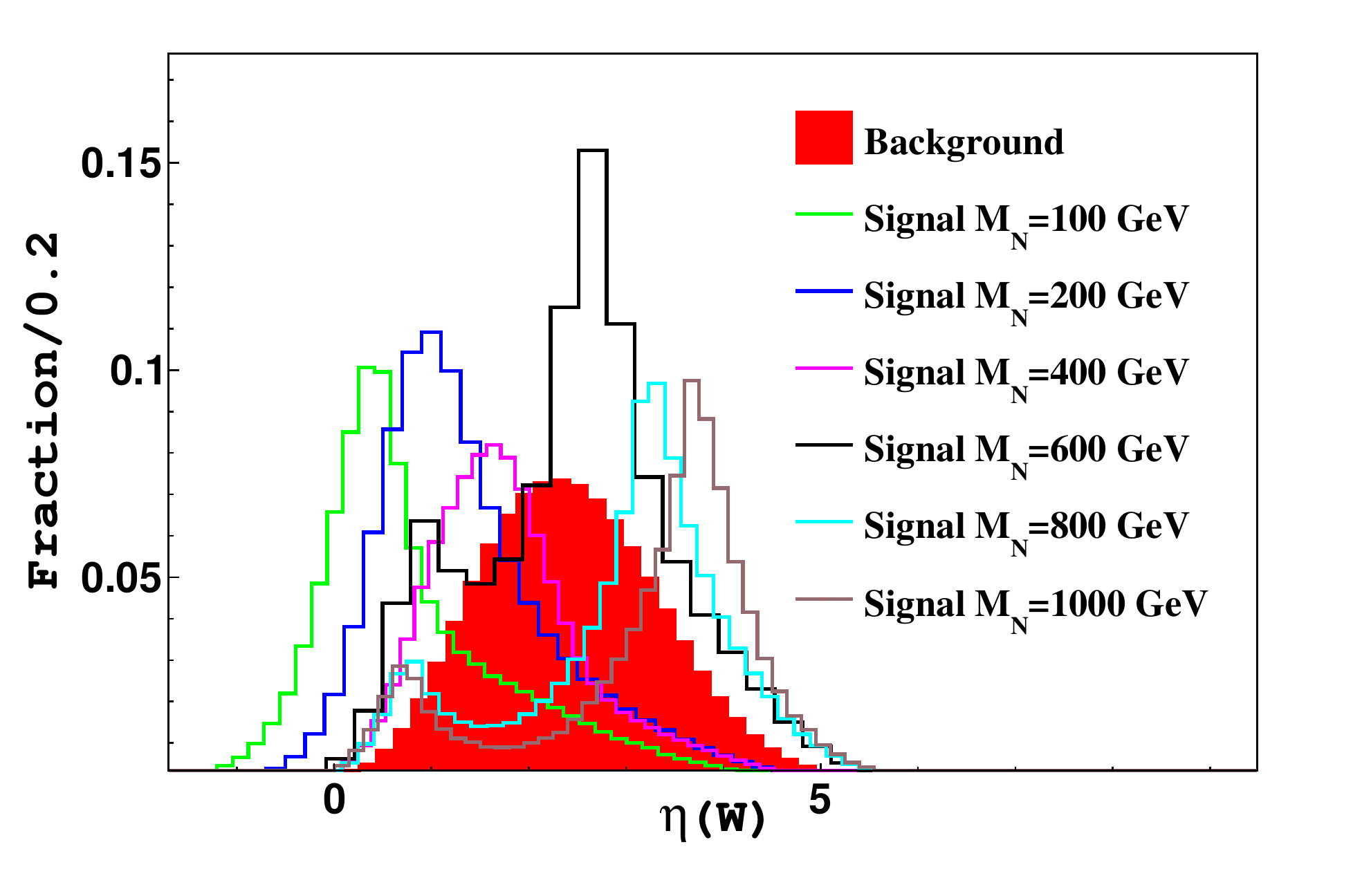}
\\
\includegraphics[width=0.35\textwidth]{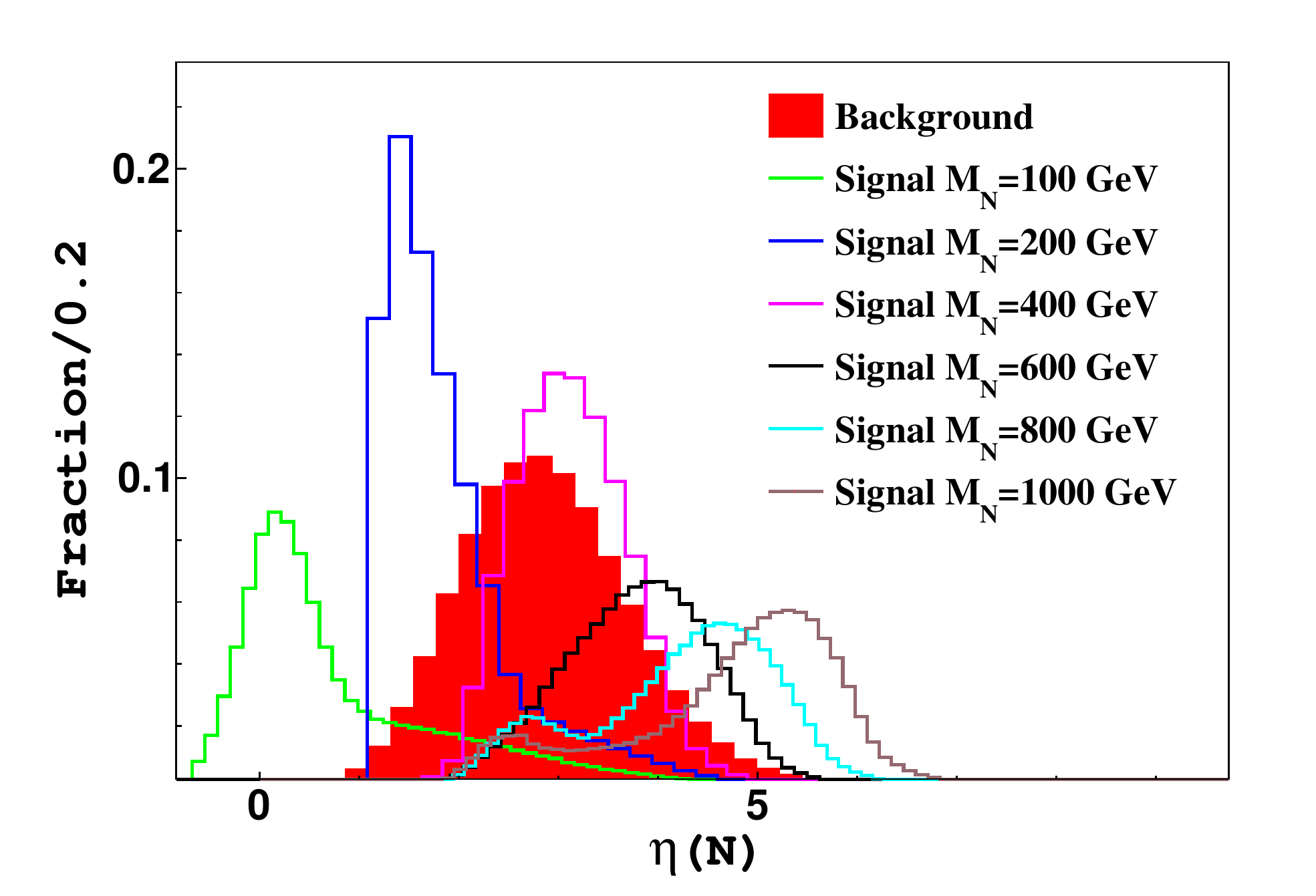}~~\includegraphics[width=0.38\textwidth]{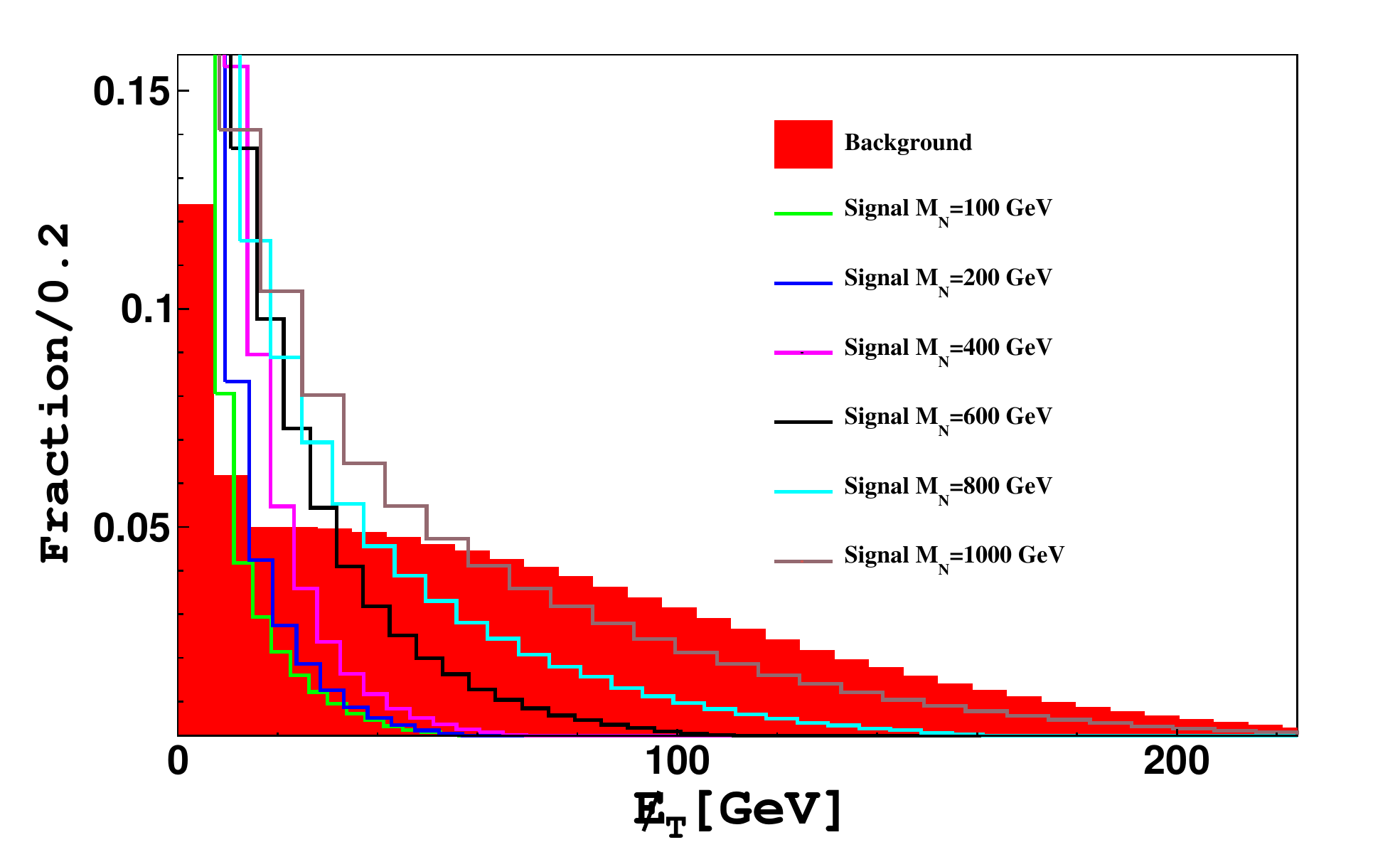}~~\includegraphics[width=0.35\textwidth]{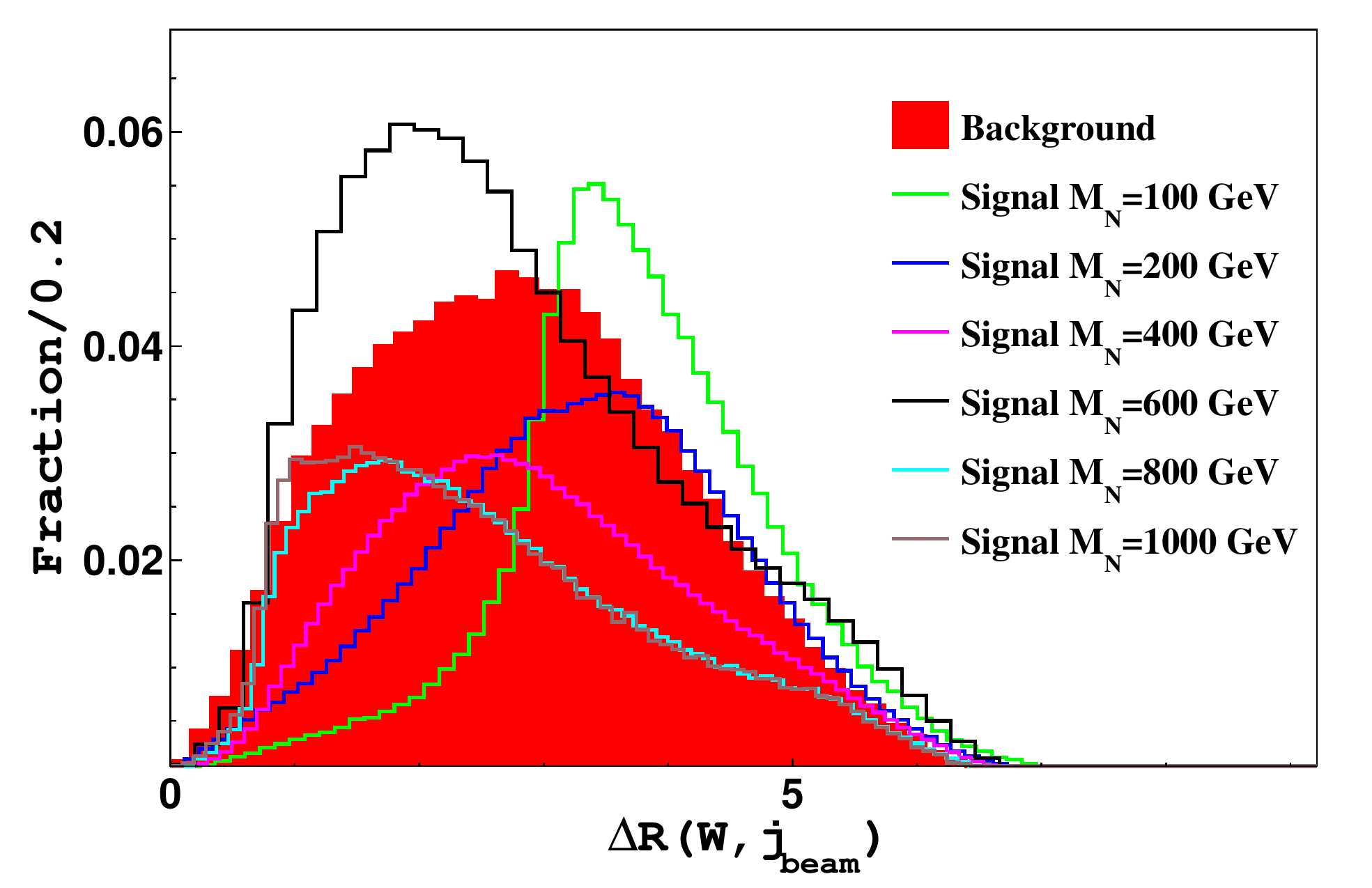}
\\
\includegraphics[width=0.38\textwidth]{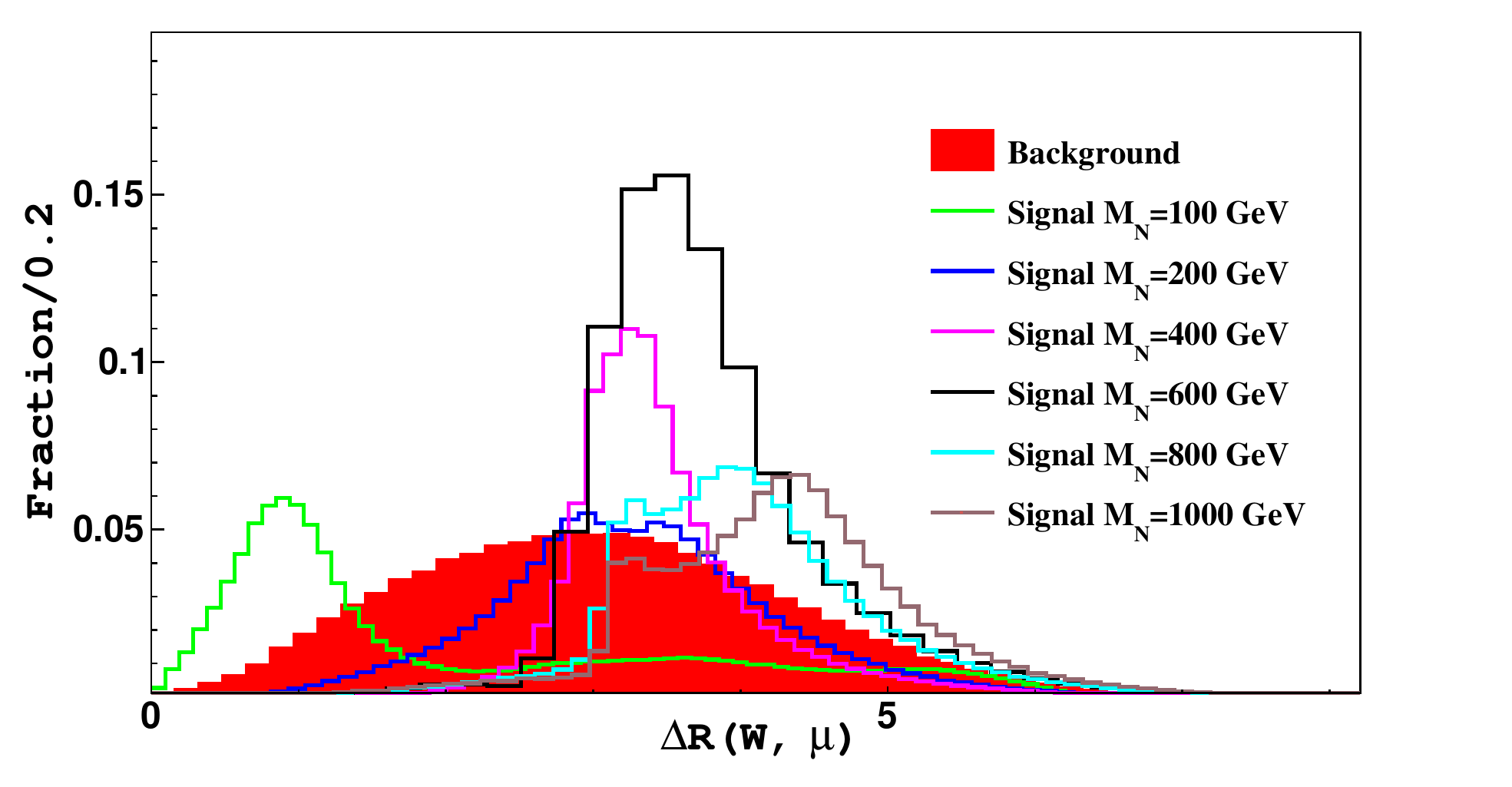}~~\includegraphics[width=0.34\textwidth]{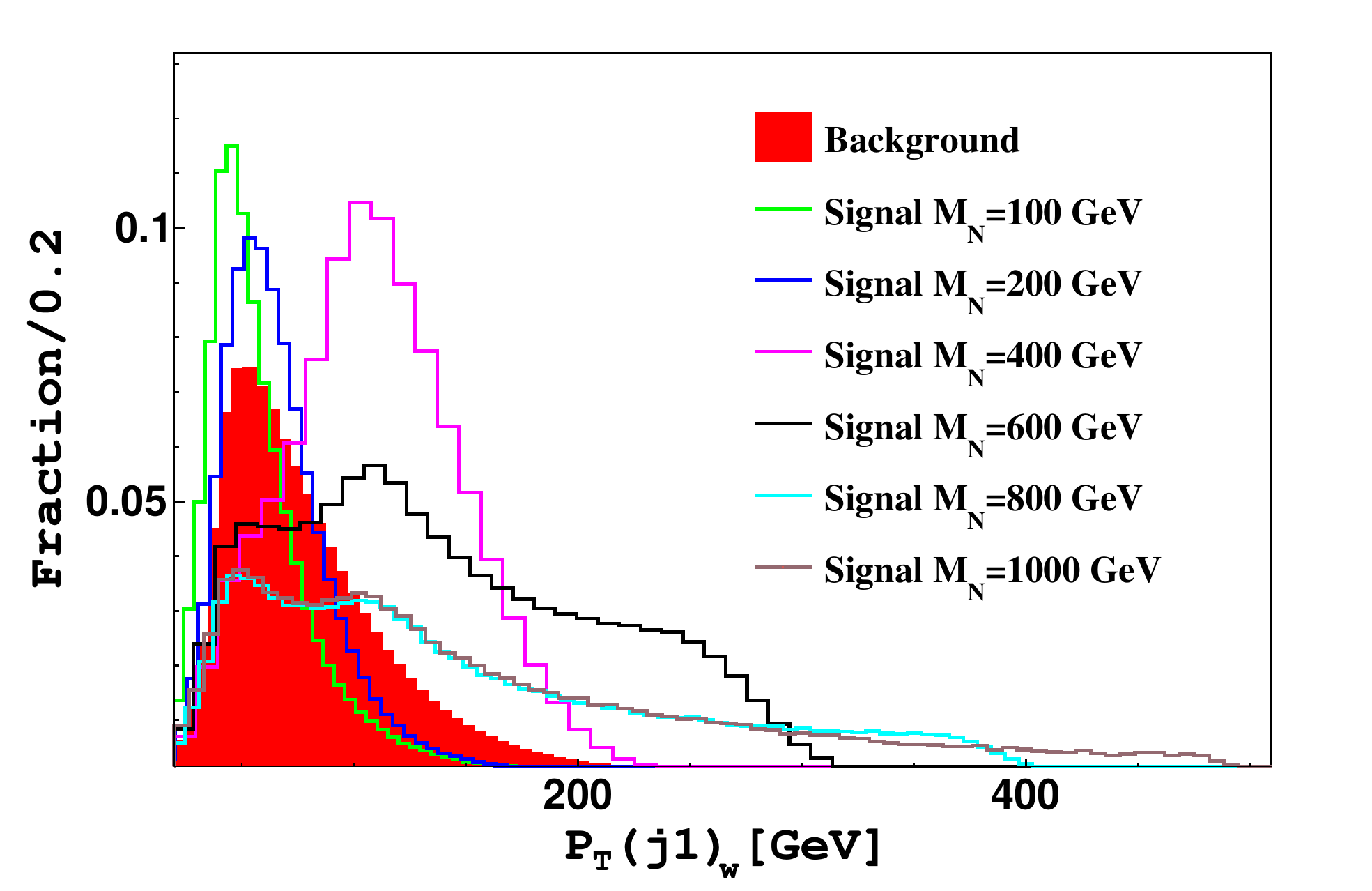}~~\includegraphics[width=0.38\textwidth]{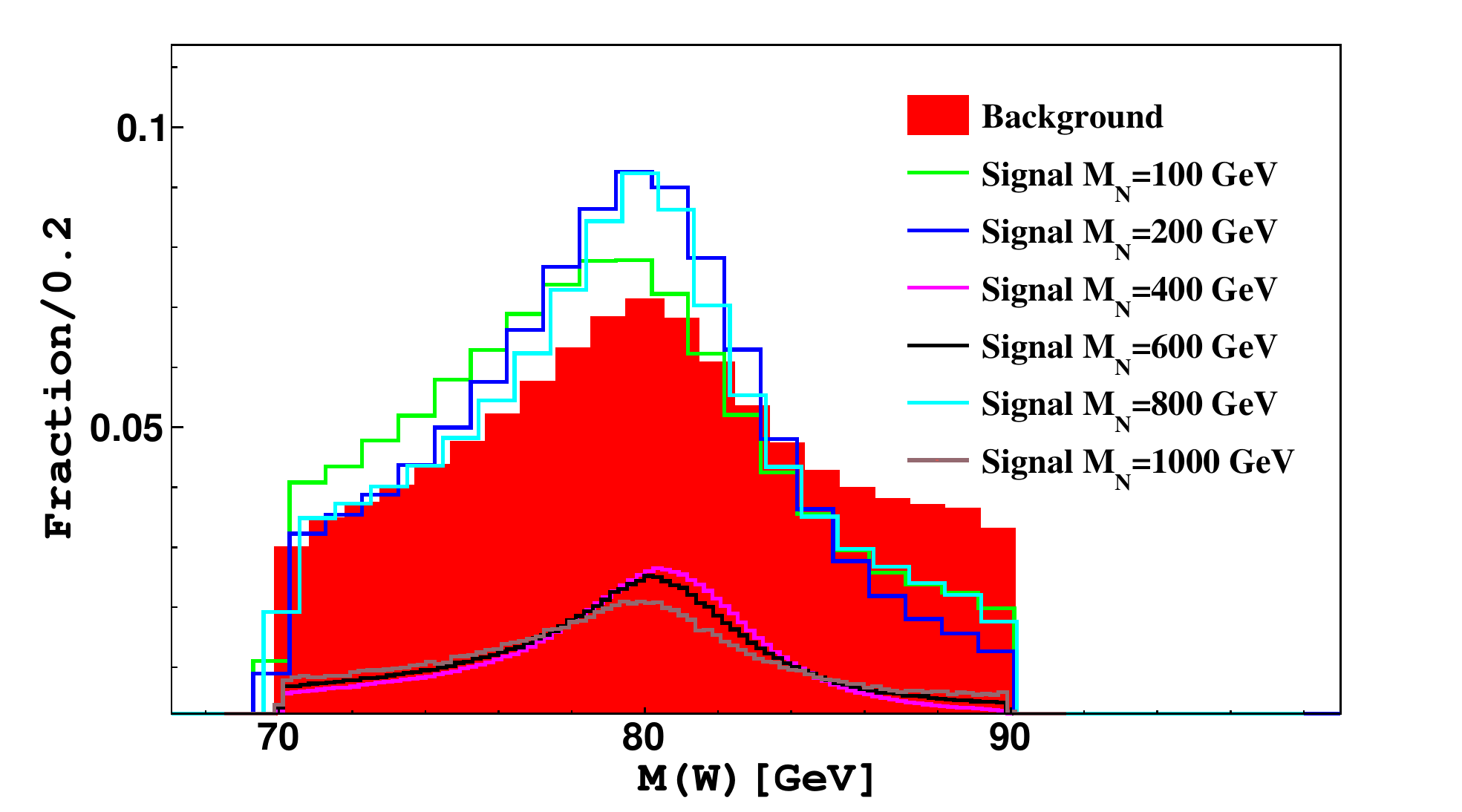}
\caption{Kinematical distributions for the $6$ signal benchmark points and all the backgrounds summed at the LHeC.}
\label{fig:input}
\end{figure}

\subsubsection{Pre-selection and analysis}
For the simulation of the signal and background event samples, the Monte Carlo event generator MadGraph5
version 2.4.3\cite{Alwall:2014hca} is employed. The parton shower and hadronisation are done by Ptyhia6\cite{Sjostrand:2006za}. For fast detector simulation we use Delphes\cite{deFavereau:2013fsa}. 
We note that Pythia needs to be patched\cite{UtaKlein} in order to achieve a reasonable event generation
efficiency and that it is crucial that the first (second) beam, as inputted in the MadGraph run card, corresponds to the proton (electron) to correctly match the asymmetric detector setup implemented in the Delphes card. 

For signal reconstruction (at reconstruction level after detector simulation) we require at least one muon with $P_T \ge 2$ GeV and three jets with $P_T \ge 5$ GeV. 
We reconstruct the $W$ boson from the possible combinations of the three jets and adopt a mass window of $ 60 \le m_W \le 100$ GeV. This allows to fix the beam jet via the one with the highest pseudo rapidity and highest momentum remaining. 
We construct $18$ kinematical distributions as input to the package TMVA\cite{TMVA2007} to perform the Multi-Variate Analysis, employing a Boosted Decision Tree (BDT). The discriminating power of the BDT relies on the fact that the signal and the background may be characterized by different features that can be entangled. 

The BDT algorithm ranks the input variables according to its ability to separate between signal events and background events. To illustrate the results we show, for the LHeC, the $18$ variable distributions for the $6$ signal benchmark points and all backgrounds summed in Fig.\ref{fig:input}. The invariant mass distribution of the heavy neutrino is classified as the highest ranking for all mass points with $m_N\ge 200$ GeV, while for smaller mass transverse missing energy is the most important one. Other variables like muon transverse momentum, $P_T (\mu)$, and the distance between heavy neutrino and the beam jet, $\Delta R(N,j_{beam})$, have high rank in separating the signal events from background events especially for $m_N\ge 400$ GeV. All the other variables have fluctuating rank according to the different mass points. 

We remark that the large asymmetry in the beam energies at an electron-proton collider leads to a strong boost of the final states particles into the direction of the proton beam, which in general shifts the angular observables towards larger $\eta$ values and affects the angular correlations.
This effect is kinematically fixed for the known SM background processes, but shows an interesting dependency on the heavy neutrino mass for the signal process.
In particular for masses of a few hundred GeV the jets from the decay chain $N \to W^+ \mu^- \to jj \mu^- $ feature a peak at large $\eta$ values, while for small masses of a few tens of GeV the heavy neutrinos and their decay products tend to reside in the backward direction at small negative $\eta$ values, cf.\ fig.\ \ref{fig:kinematics}.

\subsubsection{Results}
\begin{figure}[t]
\includegraphics[width=0.49\textwidth]{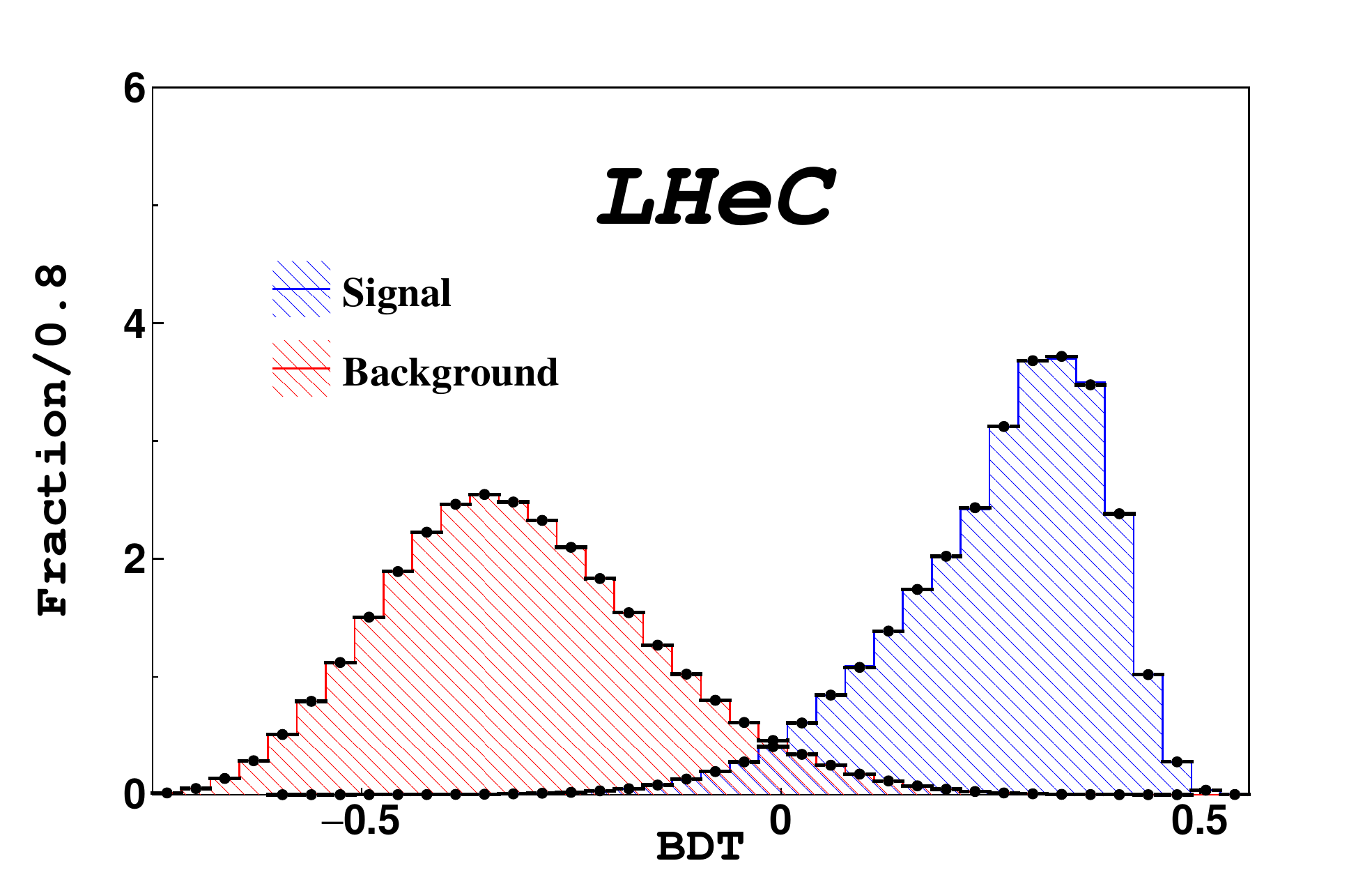}
\includegraphics[width=0.49\textwidth]{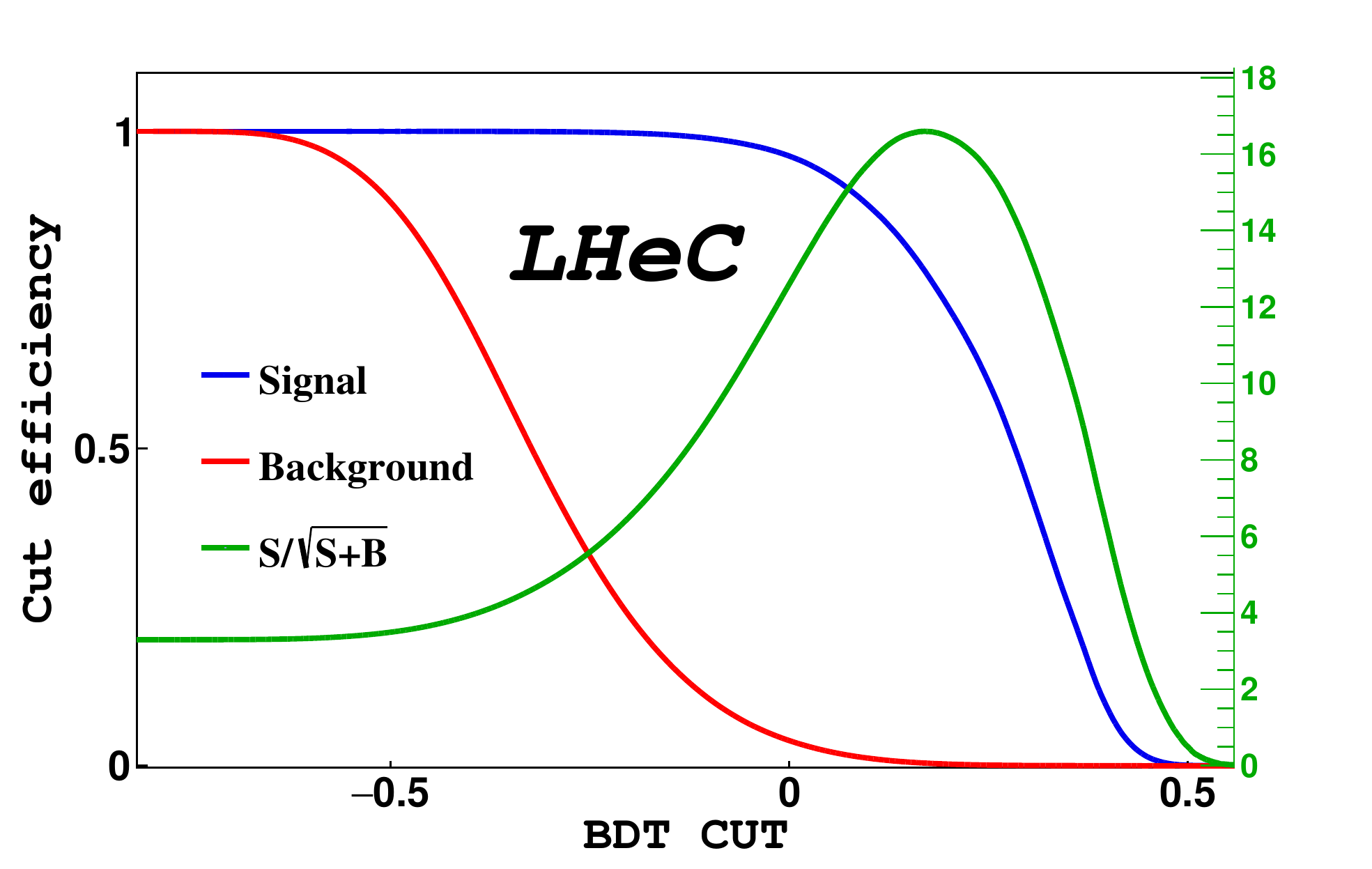}

\includegraphics[width=0.49\textwidth]{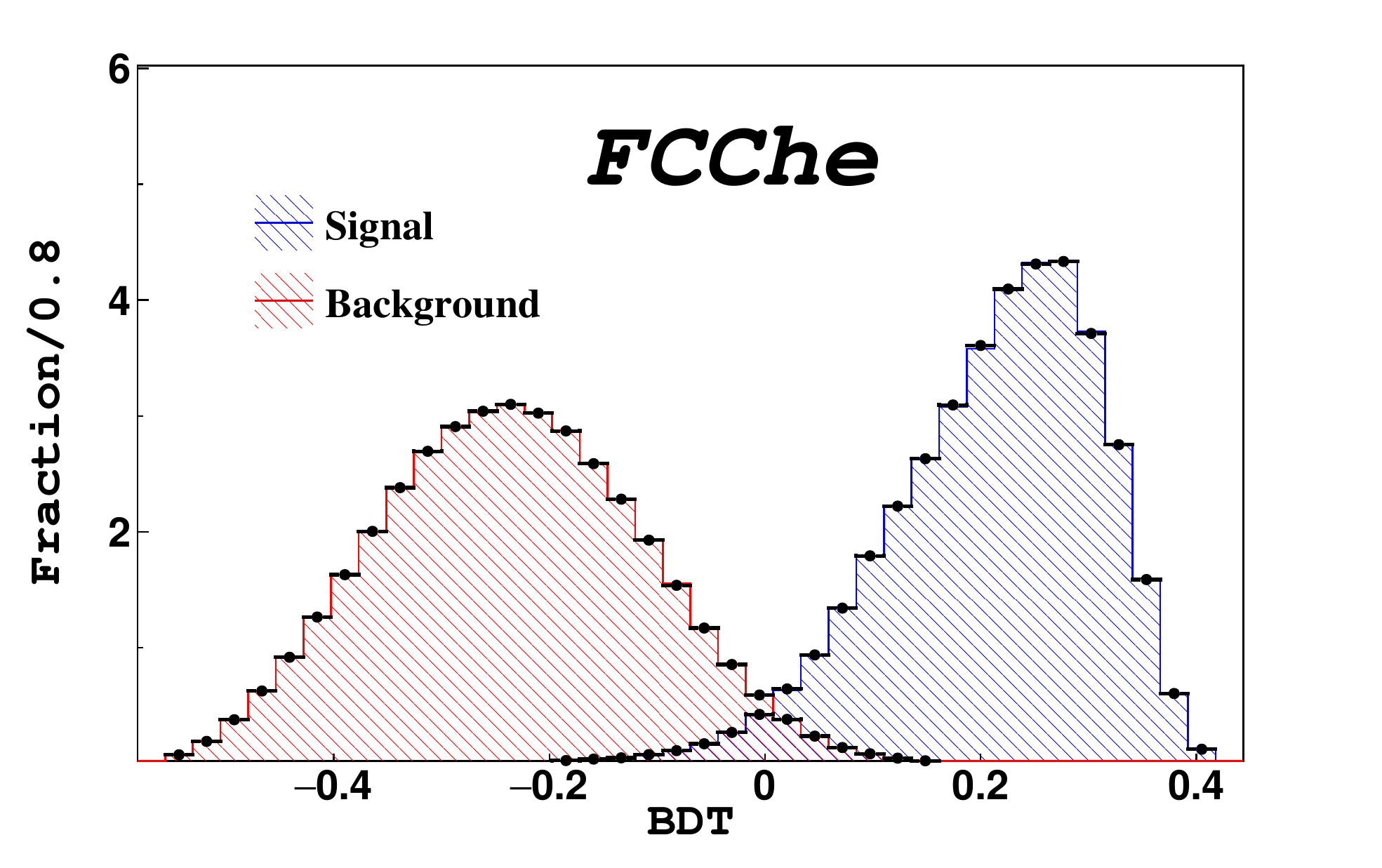}
\includegraphics[width=0.49\textwidth]{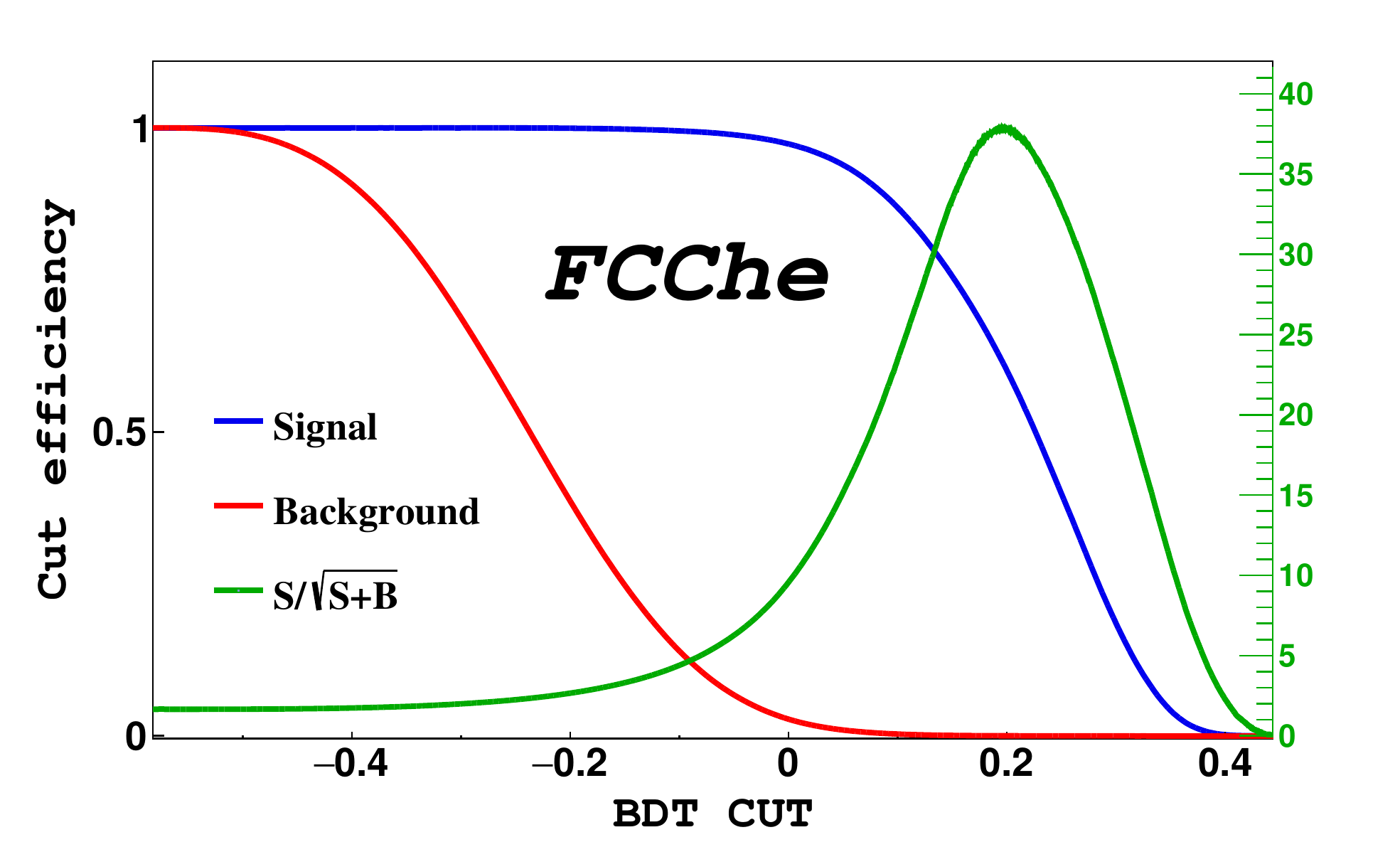}
\caption{All the plots use $M_N = 400$ GeV and $ \theta_e=\theta_\mu = 0.01$, $|\theta_\tau |= 0$.
{\it Upper left}: BDT distribution at the LHeC for both train and test samples superimposed. 
{\it Bottom left}: BDT distribution at the FCC-he for both train (black dotted distributions) and test (filled blue and red distributions) samples superimposed for both signal and background events. 
{\it Upper right}: Cut efficiency at the LHeC with BDT cut $\ge 0.17$ one can get $S/\sqrt{S+B}= 16 \sigma $ with number of signal events = $330$ and background events = $64$. The cut efficiency for the signal is $0.78$ and for the background $0.004$. 
{\it Bottom right}: Cut efficiency at the FCC-he with BDT cut $\ge 0.189$ one can get $S/\sqrt{S+B}= 37.87\sigma$ with number of signal events = $1743$ and background events = $376$. The cut efficiency for the signal is $0.6$ and for the background $0.0001$.}
\label{fig:result-BDT}
\end{figure}

\begin{figure}[t]
\includegraphics[width=0.49\textwidth]{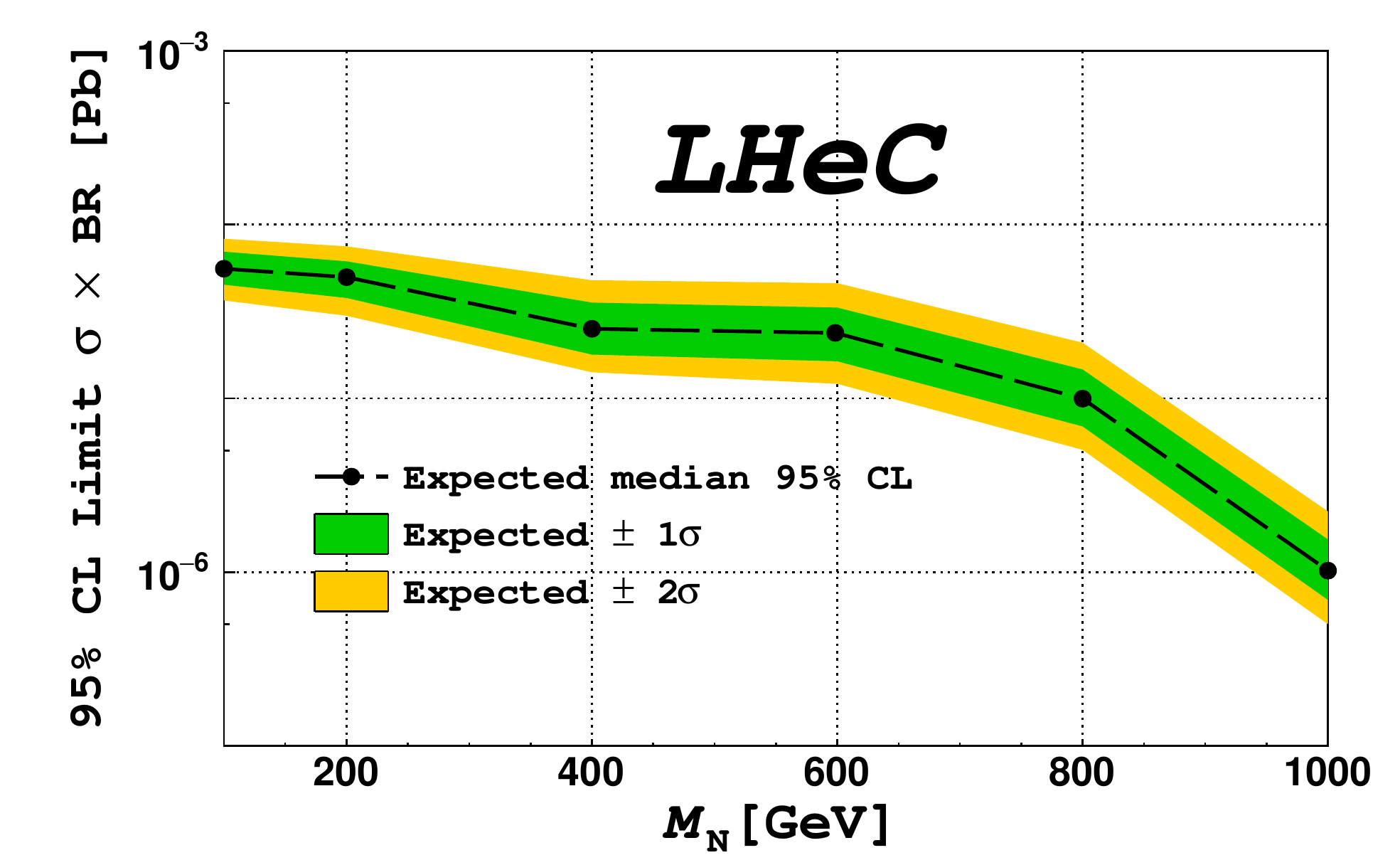}
\includegraphics[width=0.49\textwidth]{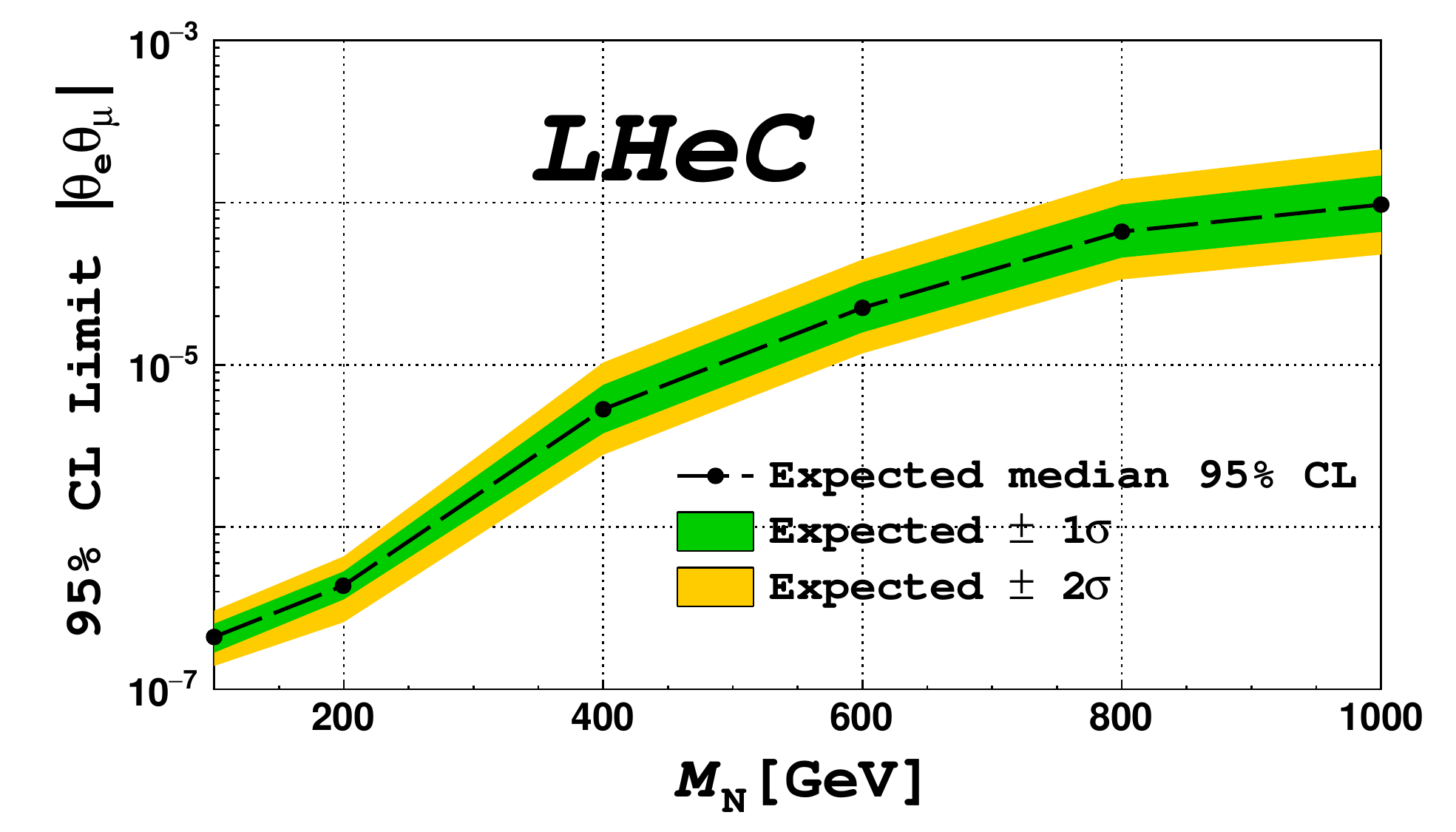}

\includegraphics[width=0.49\textwidth]{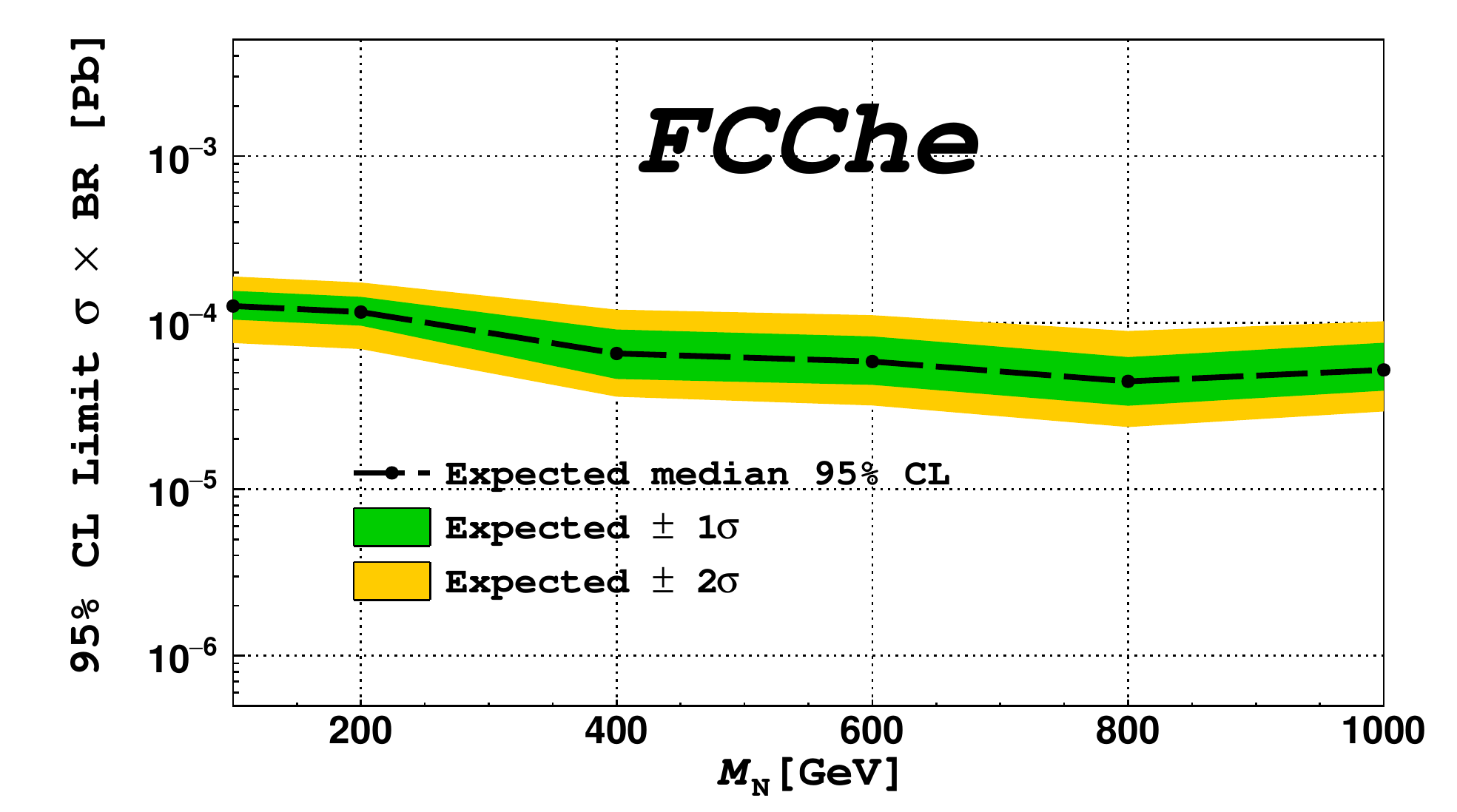}
\includegraphics[width=0.49\textwidth]{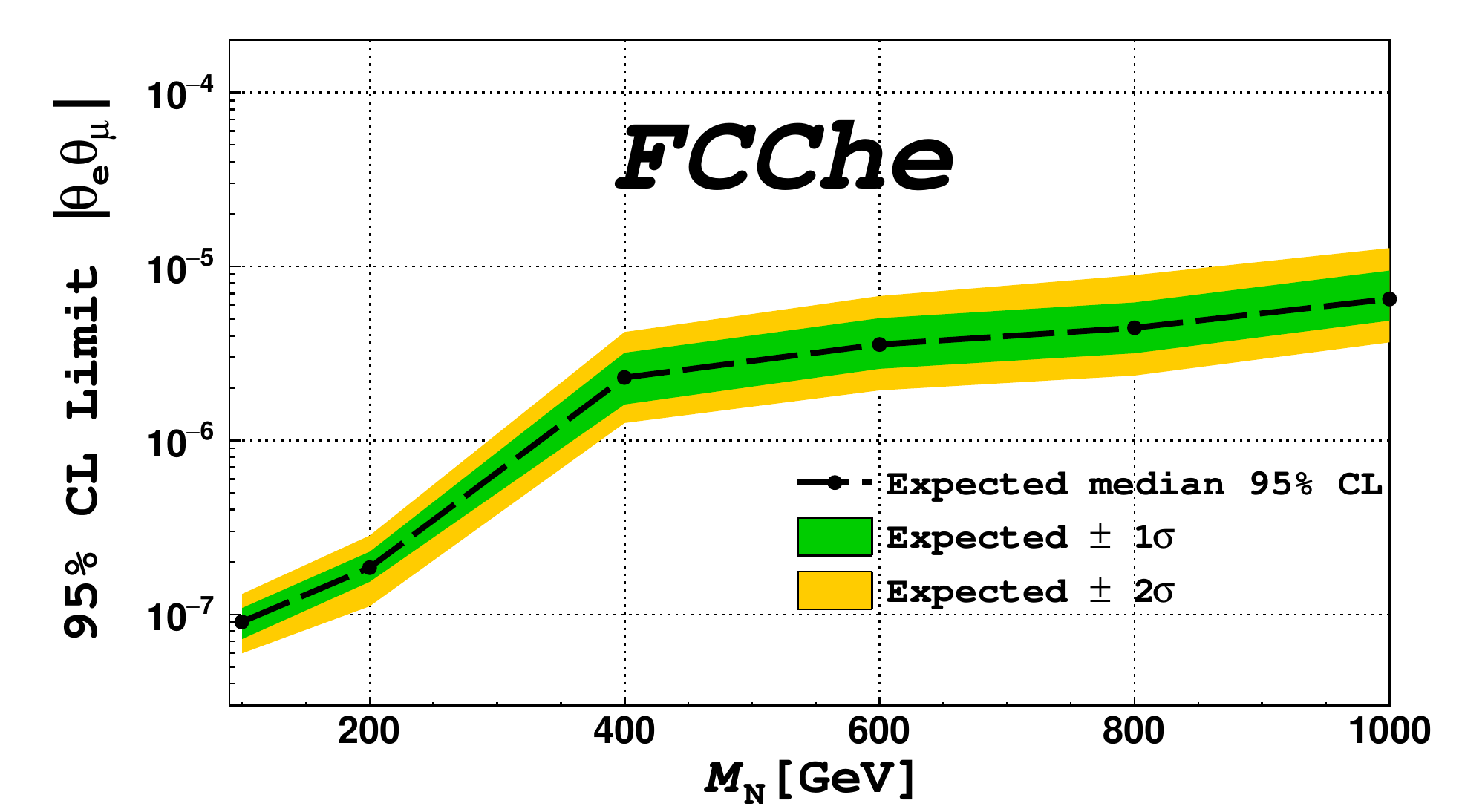}
\caption{{\it Left}: Expected limit on the production section times branching ratio of $\sigma (pe^-\to N j)\times BR(N\to \mu^- j j)$ when testing the signal hypotheses (for $|\theta_e| = |\theta_\mu |$ and $|\theta_\tau |= 0$) at LHeC (up) and FCChe(down). {\it Right}: Corresponding expected limit on the mixing parameters $|\theta_e \theta_\mu|$  when testing the signal hypotheses at the LHeC (up) and the FCChe (down).}
\label{fig:results-limits}
\end{figure}

We show the resulting BDT distributions for the LHeC and FCC-he with $m_N = 400$ GeV and $ \theta_e=\theta_\mu = 0.01$, $|\theta_\tau |= 0$
in fig.\ \ref{fig:result-BDT}. It reflects the clear separability of signal and background for moderate efficiency losses on the signal side.
The resulting limits on the cross section at 95\% confidence level are shown in the left column of fig.\ \ref{fig:results-limits}. The right column shows the resulting sensitivity on the active sterile mixing parameter combination $| \theta_e \theta_\mu |$ as a function of the heavy neutrino mass $m_N$.

Fig.~\ref{fig:result-BDT} ({\it left}) shows in blue the resulting BDT response for the LHeC and FCC-he with trained events (shown by the data points) and tested events (shown by the shaded areas) superimposed. We note that in order to avoid over-training, we require that the Kolmogorov-Smirnov classifier is around and below 0.5.
The BDT discriminator ranges from -1 to 1, events with discriminant value near 1 is classified as signal-like events (blue) and those near -1 is considered as background-like events (red). 

The optimization of signal significance as a function of signal and background cut efficiency is shown in fig.~\ref{fig:result-BDT} ({\it right}). At the LHeC, the maximum cut efficiency is at BDT $\ge 0.17$ that correspond to signal significance $\simeq 16 \sigma$ with signal efficiency 0.78 and background rejection efficiency 0.004. For the FCC-he the cut efficiency has been maximized by requiring BDT $\ge 0.189$ to obtain a signal significance $\simeq 37.8 \sigma$, with signal cut efficiency 0.6 and background rejection efficiency 0.0001. 

Based on the BDT analysis, the sensitivity for heavy neutrino searches via the lepton flavour violating process $3j+\mu$ is derived using the Higgs Analysis Combined Limit tool\cite{HiggsTool}. To extract the limits we preformed a frequentist test which uses the profile likelihood as test statistics  corresponding to the remaining number of signal/background events after the BDT cut. At the LHeC, for the benchmark point with $ \theta_e=\theta_\mu = 0.01$, $|\theta_\tau |= 0$ and $M = 400$ GeV, the number of signal events is 330 and background events 64. For the FCC-he, the number of signal events is 1743 and background events 376. 

In fig.~\ref{fig:results-limits} we show the expected median limit at $95\%$ CL with the one and two sigma bands on the total cross section ({\it left}). The right panel shows the resulting sensitivity on the related mixing angles $| \theta_e  \theta_\mu |$ (with $\theta_e=\theta_\mu $ and $|\theta_\tau |= 0$, cf.\ discussion in section 3.1). Besides the parameters of interest, such as the total cross section and the integrated luminosity, we consider an uncertainty parameter of $2\%$ for the background events as logarithmic-normal distribution to account for the unknown systematic uncertainties. Further background information on the used statistical methods can be found, e.g., in the appendix of \cite{Antusch:2018bgr}.

\subsection{Displaced vertex searches}
Heavy neutrinos with masses below the $W$ boson mass threshold and with $|\theta|^2 \leq 10^{-5}$ naturally develop lifetimes that are macroscopic, i.e.\ that allow them to travel a finite and measurable distance in the detector before they decay.
Such decays at a distance from the interaction point are reconstructed as displaced secondary vertices, which is a very exotic signature that has no irreducible SM backgrounds.
We consider the process chain $p e^- \to j (N\to {\rm visible}|_{\rm displaced})$ as our signal, where we exclude the $\sim 5\%$ branching fraction of $N \to 3 \nu$ and decays inside the detector are considered to yield unmistakable signatures.
We do not discuss here the prospects of identifying or reconstructing the heavy neutrino properties from this signature.

\subsubsection{The detector}
We use description of the LHeC detector from the CDR \cite{AbelleiraFernandez:2012cc}.
The interaction point (IP) fixes the centre of our cylindrical coordinate system, the $z$ axis is fixed by the proton beam.
The tracker has a radius of 88 cm around the $z$ axis, its z extension in forward and backward directions is about 390 cm and 190 cm, respectively.
The HCAL has a radius of 260 cm and extends an additional 217 cm and 187 cm in forward and backward direction, respectively, and the muon system adds 178 cm to the radial extension. The total detector length is 1316 cm.

\subsubsection{Vertexing}
The primary vertex can be obtained from the intersection of the charged track and the interaction region.
The interaction region has a root mean square transverse extension of $\sim 7\mu$m, and a longitudinal extension of $\sim 0.6$ mm.
The tracking resolution is $\sim 8 \mu$m. We assume that a displacement of $\sim 40 \mu$m will yield a sufficient degree of confidence that the secondary vertex is not identical to the primary vertex \cite{AbelleiraFernandez:2012cc}.
We emphasize that the considered displacement is not confined to the transverse plane since the precision of the primary vertex is known with ${\cal O}(10)\mu$m in all directions.
Thus, the minimal vertex displacement is given by $40\mu$m and the maximal vertex displacement is given by the extension of the muon system, which is 4.38 m in radial direction and 5.3 m (7.5 m) in backward (forward) direction.

\subsubsection{Backgrounds}
We discuss backgrounds only for the LHeC, the situation is very similar for the FCC-he.
Possible backgrounds come from SM particles that have a finite lifetime and are incorrectly reconstructed.
Natural candidates for such backgrounds are for instance tau leptons, which can be produced via the process $e^-p\to \nu\nu j \tau^-$ with a cross section of 
\begin{equation}
\sigma(e^- p\to \nu\nu j \tau^-) = 0.34 \text{ pb},
\end{equation}
and they have typical displacements of $\sim$ mm. 
However, tau leptons only decay either into charged leptons plus neutrinos or into hadrons plus a neutrino and will not be easily confused with the signal signature.
We therefore assume that they can be effectively vetoed against by existing tau tags, provided that $m_N \gg m_\tau$.

Another candidate for SM backgrounds are B mesons, for which we obtain an estimate via the final states $\nu b$, $\nu \bar b$, and $\nu,j, b\bar b$, with the following cross sections:
\begin{align}
\sigma(e^-p\to\nu b) = 144 |V_{ub}|^2 \text{ pb,}\\
\sigma(e^-p\to\nu j b \bar b ) = 0.54 \text{ pb}. 
\end{align}
With $|V_{ub}|= 0.004$ \cite{Barberio:2006bi} about $\sim 10^6$ singly and doubly produced b mesons with lifetimes of $\sim 1$ ps are to be expected, most of which decay typically inside the beam pipe and within a few mm from the IP. 
The doubly produced b mesons can be vetoed against with B-tag filters and the fact that there is more hadronic activity (a second b jet) close to the IP.
A more important discriminator against all $B$ mesons is their characteristic mass around 5 GeV. We will assume that this allows for complete suppression of this background when $m_N > 5$ GeV.

One more possible background process is given by cosmic muons, which may coincide with a bunch crossing and might be misidentified as two back-to-back muons.
In the following, we assume that the cosmic muons and the above mentioned SM background can be vetoed against effectively with appropriate preselection criteria on the final state, even when the displacement is as small as $\sim 40\mu$.

\subsubsection{Analysis and results}
\begin{figure}
\centering
\includegraphics[height=0.35\textwidth]{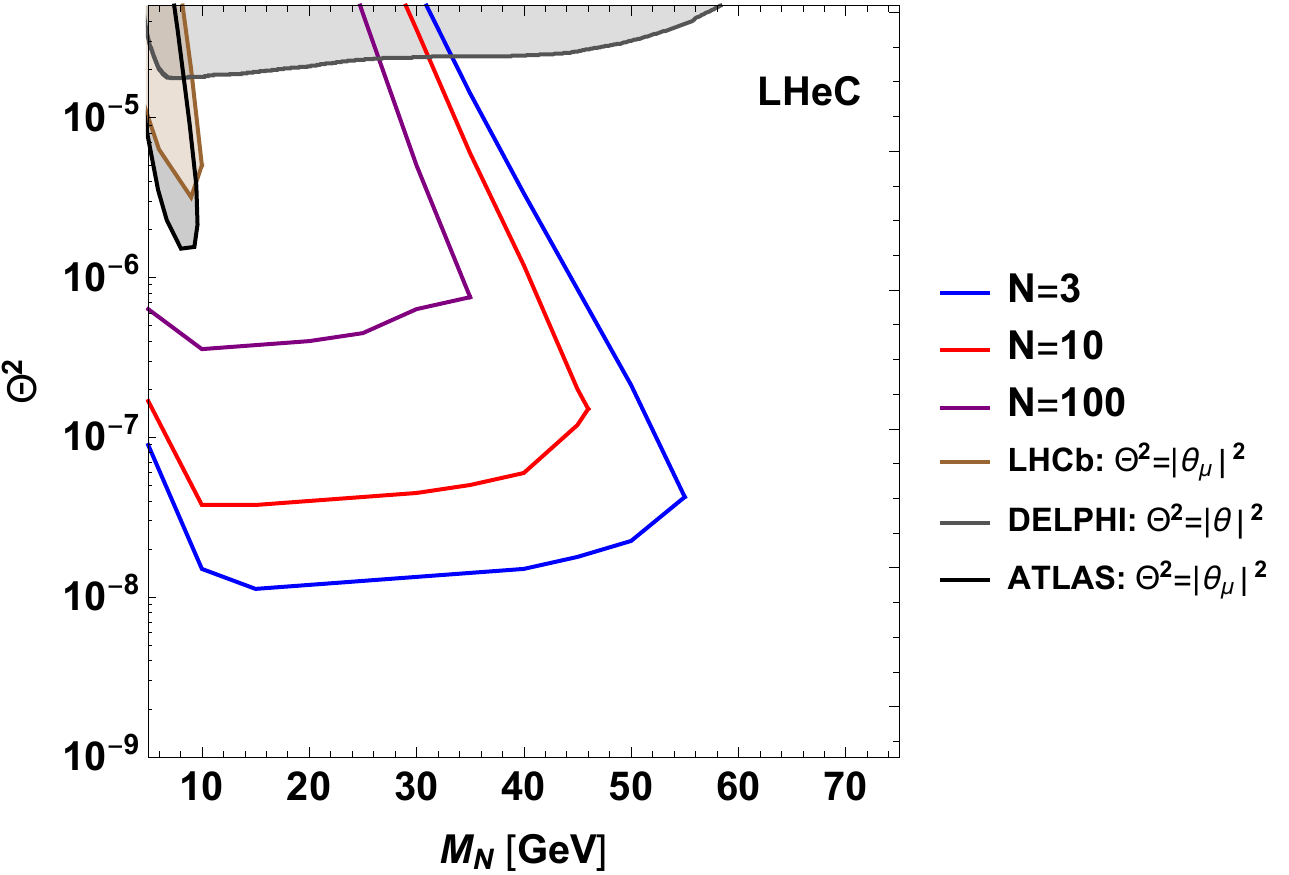}
\includegraphics[height=0.35\textwidth]{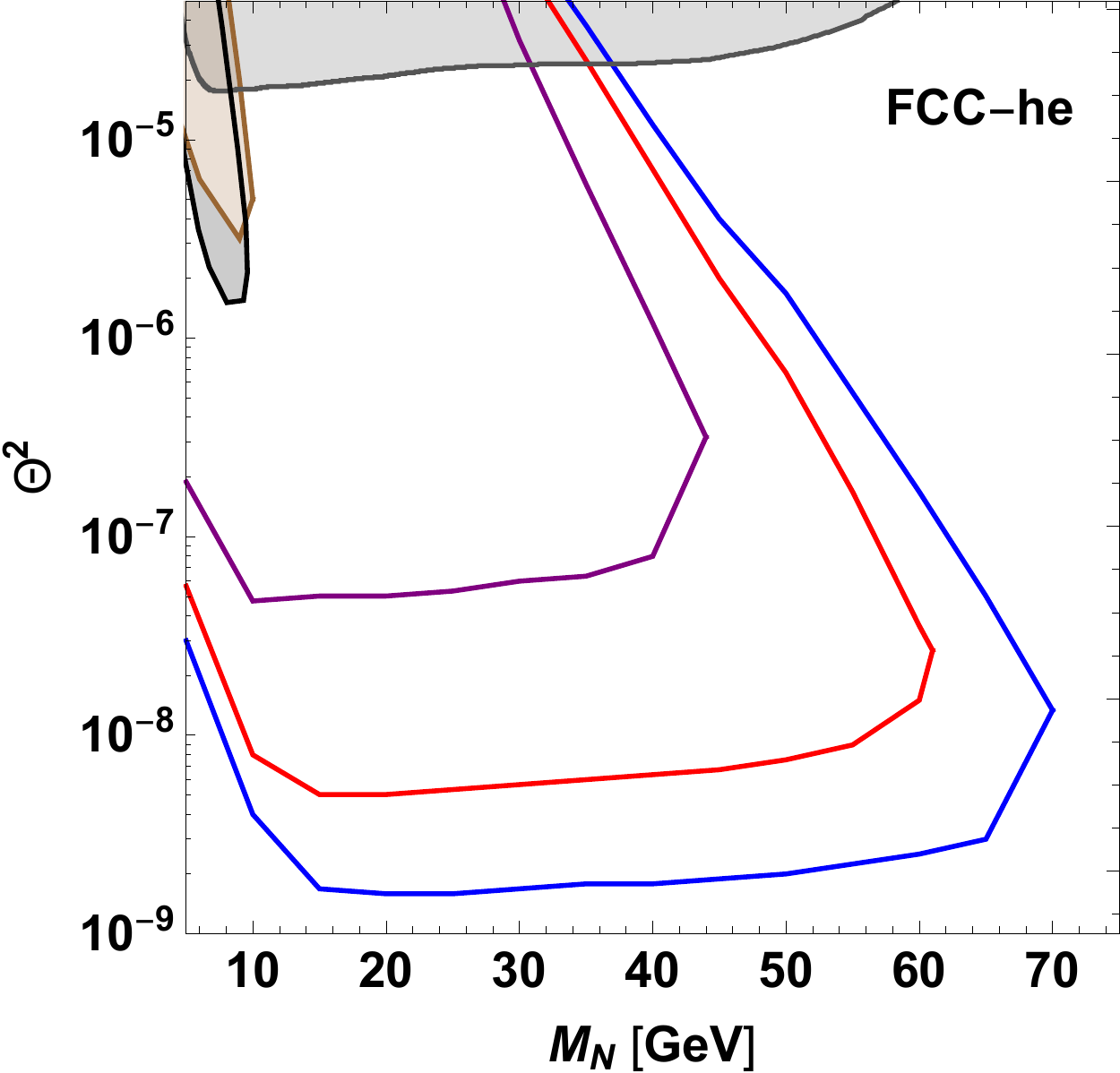}
\caption{Parameter space giving rise to $N=3,
,10,\,100$ heavy neutrino decays with a displaced secondary vertex at the LHeC (left) and the FCC-he (right). 
The gray area denotes the best exclusion limits from the experiments from ATLAS \cite{Aad:2019kiz}, LHCb \cite{Antusch:2017hhu}, LEP \cite{Abreu:1996pa}, and MEG \cite{Adam:2013mnn}. In this figure, $|\theta_\alpha|=0$ for $\alpha \neq e$.}
\label{fig:results-displaced}
\end{figure}

We quantify the expected number of heavy neutrino decays with given displacement according to the formalism presented in ref.\ \cite{Antusch:2017hhu}:
\begin{align}
&N_{\rm dv}(E_p,{\cal L},m_N,|\theta_e|) 
= 
\sigma(E_p,m_N, |\theta_e|)\, {\cal L} \times \int D_{N}(\vartheta,\gamma)\,P_{\rm dv}(x_{\rm min}(\vartheta),x_{\rm max}(\vartheta),\Delta x_{\rm lab}(\uptau,\gamma))\, d\vartheta d\gamma\,.
\label{eq:masterequation}
\end{align}
In the above equation, $\sigma$ labels the production cross section and depends and the proton beam energy $E_p$,  ${\cal L}$ the integrated luminosity, $D_{N}(\vartheta,\gamma)$ is the probability distribution for $N$ with an angle $\vartheta$ between momentum $p$ and beam axis, $P_{\rm dv}$ is the probability distribution of a decay, and $\uptau$ is the proper life time.
The probability of decays with a displacement $x_{\rm min} \leq \Delta x_{\rm lab} \leq x_{\rm max}$ is 
\begin{equation}
P_{\rm dv}= {\rm Exp}\left(\frac{-x_{\rm min}}{\Delta x_{\rm lab}}\right) -  {\rm Exp}\left(\frac{-x_{\rm max}}{\Delta x_{\rm lab}}\right)\,.
\label{eq:probability}
\end{equation}
We take the asymmetric set up of the detector and the full angular and momentum distributions into account and choose for our analysis the 95\% confidence level, corresponding to the number of displaced vertices being $N_{\rm dv} \geq 3.09$.
We show the corresponding exclusion sensitivity contour at 95\% confidence, labelled ``N = 3'' in fig.\ \ref{fig:results-displaced}. The figure also contains the contour lines for the number of expected displaced vertices being $N=10,\,100$ for comparison.
It is worth noticing that most of the decays enclosed inside the contour yield events in the backward hemisphere of the detector, i.e.\ into the direction of the electron beam, where there is indeed no background to be expected, cf.\ fig.\ \ref{fig:kinematics}.

\subsection{Discussion}
\begin{figure}
\centering
\includegraphics[width=0.7\textwidth]{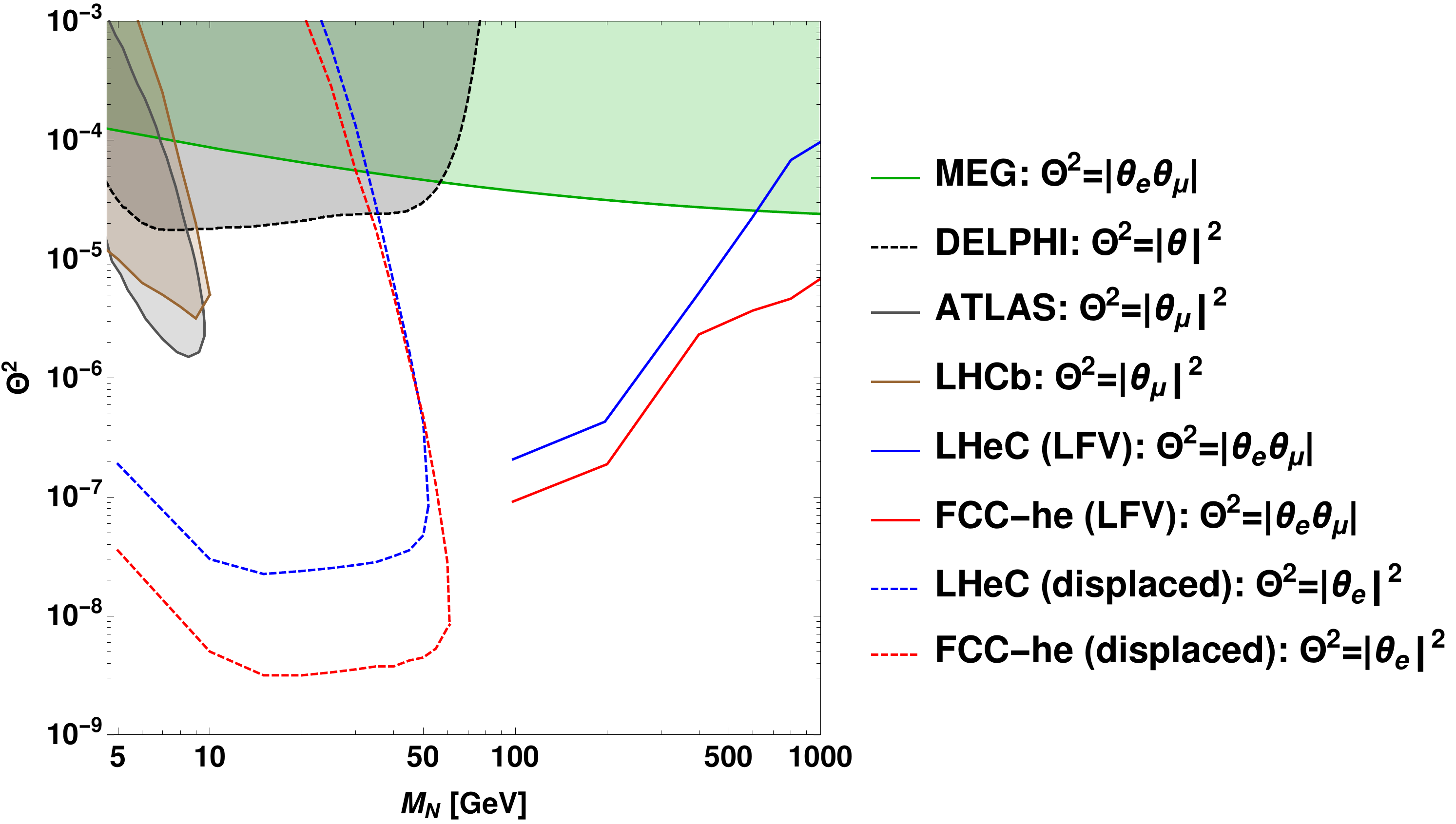}
\caption{Sensitivity of the LFV lepton-trijet searches (at 95\% C.L.) and the displaced vertex searches (at 95\% C.L.) compared to the current exclusion limits from ATLAS \cite{Aad:2019kiz}, LHCb \cite{Antusch:2017hhu}, LEP \cite{Abreu:1996pa}, and MEG \cite{Adam:2013mnn}. The sensitivity of the lepton-trijet searches at ep colliders can be generalized to its full $\theta_\alpha$-dependence by replacing $| \theta_e \theta_\mu |$ with $2 |\theta_e|^2 |\theta_\mu|^2/|\theta|^2$.}
\label{fig:complementarity}
\end{figure}

To put our results into a general context we show a combination of the leading search prospects for heavy neutrinos (at 95\% confidence level) in fig.\ \ref{fig:complementarity}, together with present constraints from the MEG experiment \cite{Adam:2013mnn} as computed in ref.\ \cite{Antusch:2015mia}, the LEP experiment Delphi \cite{Abreu:1996pa}, and the recent result from ATLAS \cite{Aad:2019kiz}.
In comparison, the searches for lepton number conserving and lepton flavor violating final states at the HL-LHC and FCC-hh can only test active-sterile mixings that are much larger \cite{Antusch:2018bgr}.

It is worthwhile to compare the results of this study with previous ones from ref.\ \cite{Antusch:2016ejd}, which was a first look at the parton level and considered only a single background process.
Here we performed an analysis at the reconstructed level, including hadronization and a number of backgrounds, and the obtained results -- optimised by the use of the BDT -- are more robust than the previous ones.
It turns out that the new results have a better sensitivity to active-sterile mixing for any given mass compared to the previous result; as an illustration, the sensitivity for $M_N = 200$ GeV at 1$\sigma$ parton level was $2\times 10^{-6}$ while here it is $2\times 10^{-7}$ at 95\% confidence. 
This is because the previous results were on purpose very conservative in employing only a single cut on the missing energy.
In this light it would be very interesting to compare our results with other promising signatures in ref.\ \cite{Antusch:2016ejd}, such as the dilepton-jet final state in the high-mass regime or some of the signatures from $W\gamma$ fusion at high energies.

Let us comment on the impact of the flavor structure of active-sterile mixing.
In the scenario that is complementary to our choice above, where $ |\theta_\mu| \ll |\theta_e|,|\theta_\tau|$, the LFV final state $\tau^- jjj$ is the most prominent.
We expect that our results are indicative also for this case because the tau reconstruction should benefit from the clean and pile up-free environment of the electron-proton collision, such that the reduction of the signal efficiency due to reconstruction losses should be small. 
Therefore our results should hold in more generality, unless unless $|\theta_e| \gg |\theta_\mu|,|\theta_\tau|$, in which case the lepton flavor conserving signatures become most relevant.

While in our model lepton number violation (LNV) is effectively absent for masses of ${\cal O}(100)$ GeV \cite{Antusch:2017ebe} it is interesting to consider the possibility of the LNV final states at lower masses.
The lepton trijet signature with an anti-lepton is also free of background and can be detectable with a significance that is similar to the lepton number conserving lepton trijet.
This is important for the investigation of heavy neutrino-antineutrino oscillations, a phenomenon that can arise naturally in our model when the heavy neutrino pair is almost mass-degenerate and has macroscopic lifetimes. 
This phenomenon is rooted in the interference between the two Majorana-like heavy neutrinos and suppresses or allows lepton number violation (LNV) as a function of the displacement of the secondary vertex (or more precisely of the heavy neutrino lifetime).
When the heavy neutrino production and decay vertices are separable in the detector these oscillations can be observed experimentally at ep colliders via the unambiguous LNV signature $N \to \ell^+ J$, where $J$ denotes a number of hadrons.
Therefore this signature could be observable at the LHeC and FCC-he within the contour lines shown in fig.\ \ref{fig:results-displaced} and with sufficient statistics even a determination of the oscillation length could be possible, which allows for instance to infer the mass splitting and thereby contribute to testing the conditions for leptogenesis.

\section{Conclusions}

In this paper we have analysed two of the most promising signatures of heavy neutrinos at ep colliders: the lepton-flavour violating (LFV) lepton-trijet signature $p\ e^- \to  \mu^- + \, 3 j$ and the displaced vertex signature. The latter is particularly relevant for heavy neutrino masses below $m_W$, where the heavy neutrinos can have macroscopic lifetimes. The lepton-trijet signature has been identified e.g.\ in ref.\ \cite{Antusch:2016ejd} as one of the most promising signatures among the many possible search channels for all collider types in the mass region above $m_W$ up to some hundreds of GeV. 

To capture the heavy neutrino properties of low scale seesaw models, we have used the ``Symmetry Protected Seesaw Scenario'' (SPSS) benchmark model \cite{Antusch:2015mia}, which includes two sterile neutrinos with opposite charges under  a ``lepton number''-like symmetry.  
We have performed our analysis for the choice $\theta_e=\theta_\mu $ and $\theta_\tau = 0$ for the active-sterile mixing angles. However, e.g.\ for the lepton-trijet signature, replacing $| \theta_e \theta_\mu | $ by $ 2 |\theta_e|^2 |\theta_\mu|^2/|\theta|^2$, one can easily recover the full parameter dependence.
 
We also note that we have used the ``symmetry limit'' of the benchmark model for our analysis, such that all final states are lepton number conserving. When the light neutrino masses are introduced via a small breaking of the protective symmetry, this can in principle (depending on the induced small mass splitting of the quasi-degenerate heavy neutrino pair) lead to observable lepton number violation via heavy neutrino-antineutrino oscillations. For displaced vertices the heavy neutrino-antineutrino oscillations
might even be resolved via an oscillatory lifetime-dependence of $ \mbox{Br}(N\to \mu^- + \, 2 j) /  \mbox{Br}(N\to \mu^+ + \, 2 j )$, as discussed in \cite{Antusch:2017ebe}. 

Regarding the displaced vertex signatures, we have improved previous estimates by including the full detector geometry and the distribution of the relativistic velocity of the heavy neutrinos. We found that LHeC and FCC-he can reach remarkable exclusion sensitivities down to ${\cal O}(10^{-8})$ and ${\cal O}(10^{-9})$ for $| \theta_e  \theta_\mu |$, respectively, at the 95\% confidence level (cf.\ figs.\ \ref{fig:results-displaced} and \ref{fig:complementarity}) . 
For the LFV lepton-trijet signature at ep colliders, we improved on previous estimates by including SM background processes and separating signal from background signatures at the reconstructed level with a Boosted Decision Tree (BDT).
Our statistical evaluation shows that this channel can reach exclusion sensitivities to active-sterile mixing parameters $| \theta_e  \theta_\mu |$  as small as $10^{-7}$ for FCC-he and $2\times 10^{-7}$ for LHeC at the 95\% confidence level. 
For the considered benchmark model, this is the best sensitivity of all currently discussed heavy neutrino signatures in this mass range. For the whole mass region between about $5$ GeV and up to ${\cal O}$(1 TeV) the sensitivity prospects for these signatures are reaching deeply into the currently unconstrained region.

In summary, our results demonstrate that ep colliders, such as the LHeC and the FCC-he, are excellent facilities for discovering heavy neutrinos in a large mass window around the electroweak scale. They are particularly good in the mass region above $m_W$ up to some hundreds of GeV, where the LFV lepton-trijet signature could be a ``golden channel'' for heavy neutrino searches. A discovery of heavy neutrinos would have far-reaching consequences, opening up the possibility to resolve the origin of the observed neutrino masses, which is one of the great open questions in particle physics.

\subsection*{Acknowledgements}
We thank Max and Uta Klein, Monica D'Onofrio, and Georges Azuelos for useful discussions. 
This work has been supported by the Swiss National Science Foundation.
O.F.\ received funding from the European Unions Horizon 2020 research and innovation program under the Marie Sklodowska-Curie grant agreement No 674896 (Elusives).

%

\bibliographystyle{unsrt}

\end{document}